\documentclass[traditabstract]{aa}
\usepackage{txfonts}
\usepackage{graphicx}
\usepackage{rotating}
\usepackage{natbib}
\bibpunct{(}{)}{;}{a}{}{,}

\newcommand{\ie}{i.e. }
\newcommand{\eg}{e.g. }

\begin{document}

\title{Cosmological parameter extraction and biases from Type Ia supernova magnitude evolution}
\author{Sebastian Linden\inst{1} \and Jean-Marc Virey\inst{1} \and Andr\'e Tilquin\inst{2}} 
\institute{Centre de Physique Th\'eorique\thanks{Centre de Physique Th\'eorique is UMR 6207 - `Unit\'e Mixte de Recherche' of CNRS and of the Universities `de Provence', `de la Mediterran\'ee', and `du Sud Toulon-Var' - Laboratory affiliated with FRUMAM (FR2291).}, Universit\'e de Provence, CNRS de Luminy case 907, F-13288 Marseille cedex 9, France. E-Mail: linden@cpt.univ-mrs.fr. \and Centre de Physique des Particules de Marseille, Universit\'e de la Mediterran\'ee, CNRS de Luminy case 907, F-13288 Marseille cedex 9, France.}
\abstract{We study different one-parametric models of type Ia supernova magnitude evolution on cosmic time scales.\\
Constraints on cosmological and supernova evolution parameters are obtained by combined fits on the actual data coming from supernovae, the cosmic microwave background, and baryonic acoustic oscillations. We find that the best-fit values imply supernova magnitude evolution such that high-redshift supernovae appear some percent brighter than would be expected in a standard cosmos with a dark energy component. However, the errors on the evolution parameters are of the same order, and data are consistent with nonevolving magnitudes at the $1\sigma$ level, except for special cases.\\
We simulate a future data scenario where SN magnitude evolution is allowed for, and neglect the possibility of such an evolution in the fit. We find the fiducial models for which the wrong model assumption of nonevolving SN magnitude is not detectable, and for which biases on the fitted cosmological parameters are introduced at the same time. Of the cosmological parameters, the overall mass density $\Omega_{\mathrm{M}}$ has the strongest chances to be biased due to the wrong model assumption. Whereas early-epoch models with a magnitude offset $\Delta m\sim z^2$ show up to be not too dangerous when neglected in the fitting procedure, late epoch models with $\Delta m\sim\sqrt{z}$ have high chances of undetectably biasing the fit results.
}
\maketitle

\section{Introduction}
An important cosmological probe and the historically first indicator of a presently accelerating expansion of the universe are type Ia supernovae (SNe Ia, see \citet{Riess98,P99}). SNe Ia are rare, massively luminous cosmic events that are believed to be violent thermonuclear explosions of degenerate white dwarfs, resulting from the ignition of degenerate nuclear fuel in stellar material \citep{Hoyle60}. Observations of nearby extragalactic SNe Ia indicate that they could be standardized, and could therefore be used as a distance measure. In a universe described by a Robertson-Walker metric,
\begin{equation}
ds^2=-c^2d\tau^2+a^2(\tau)\left[d\chi^2+f^2(\chi)(d\theta^2+(\sin\theta)^2d\phi^2)\right],
\end{equation}
cosmic events of intrinsic luminosity $\mathcal{L}$ would yield a measured flux,
\begin{equation}
\mathcal{F}=\frac{\mathcal{L}}{4\pi(1+z)^2a_0^{\;2}f^2(\chi)},
\label{eq:fluxF}
\end{equation}
on Earth, where $a_0=a\left(\tau_0\right)$ is the cosmological scale factor at the moment of reception (0), $\chi=\chi_{\mathrm{e}}-\chi_0$ the comoving distance between the emittor (e) and the receptor, and $f\left(x\right)=\{\sin{x},x,\sinh{x}\}$ according to a \{negative, zero, positive\} curvature index $k$ of the 3-space, respectively. Further, $z=\left(\lambda_{\mathrm{e}}-\lambda_0\right)/\lambda_{\mathrm{e}}$ is the redshift of the emitted light as observed on Earth. Equation (\ref{eq:fluxF}) accounts for the diminishing luminosity due to universal expansion of the emitted light-sphere, and to doppler-shift, where each contributes a factor $\left(1+z\right)$, and $\left(1+z\right)a_0f\left(\chi\right)$ is the luminosity distance that can be expressed as a function of redshift $z$:
\begin{equation}
d_{\mathrm{L}}(z)=\frac{c}{H_0}\frac{(1+z)}{\sqrt{|\Omega_{\mathrm{k}}|}}f\left[\sqrt{|\Omega_{\mathrm{k}}|} \int\limits_0^z { \frac{\mathrm{d}\tilde{z}}{E(\tilde{z})} }\right],
\end{equation}
where $\Omega_{\mathrm{k}}=-kc^2/\left(a_0H_0\right)^2$. Here, $H_0$ is the Hubble constant. Via the Friedmann equations of General Relativity the Hubble function $H\left(z\right)=H_0E\left(z\right)$ can be expressed in terms of the dynamical variables of the universe's matter and energy content: \begin{eqnarray}
\nonumber E^2(z;\Omega_i,w_{\mathrm{x}})&=&\Omega_{\mathrm{k}}(1+z)^2+\Omega_{\mathrm{M}}(1+z)^3+\Omega_{\mathrm{R}}(1+z)^4+ \\
&+&\Omega_{\mathrm{x}} \mathrm{Exp} \left[ 3\int_{0}^{\ln(1+z)}\left(1+w_{\mathrm{x}}(\tilde{z})\right)\mathrm{d}\ln(1+\tilde{z})\right].
\end{eqnarray}
Therein $\Omega_i$ depicts the matter densities of baryonic and dark matter $\left(\Omega_{\mathrm{M}}=\Omega_{\mathrm{b}}+\Omega_{\mathrm{cdm}}\right)$, radiation $\left(\Omega_{\mathrm{R}}\right)$, and the dark energy component $\left(\Omega_{\mathrm{x}}\right)$, and $w_{\mathrm{x}}=p_{\mathrm{x}}/(\rho_{\mathrm{x}} c^2)$ is the possibly time varying equation of state of the dark energy component. The curvature parameter $\Omega_{\mathrm{k}}$ is determined by $\Omega_{\mathrm{k}}=1-\sum{\Omega_i}$. It is common to introduce the `Hubble free' luminosity distance $D_{\mathrm{L}}=H_0c^{-1}d_{\mathrm{L}}$ and to give $\left(c/H_0\right)$ in Mpc. Adopting the definition of apparent magnitude, $m=-2.5\log{\mathcal{F}}$, we obtain from Eq.(\ref{eq:fluxF}) the magnitude-redshift relation:
\begin{equation}
m=-2.5\log{\mathcal{F}}=M_{\mathrm{s}}+5\log{D_{\mathrm{L}}(z;\Omega_i,w_{\mathrm{x}})}.
\label{eq:magred}
\end{equation}
The normalization parameter $M_{\mathrm{s}}=2.5\log{\left[4\pi\left(c/H_0\right)^2\mathcal{L}^{-1}\right]}$ is a constant number for all sets of luminous events with the same intrinsic magnitude $M_{\mathrm{intr}}=-2.5\log{\mathcal{L}}$, and Eq.(\ref{eq:magred}) becomes a fundamental and powerful tool for extracting the cosmological parameters $\{p_i\}=\{\Omega_i,w_{\mathrm{x}}\}$ via the observables flux and redshift.\\
\indent The use of the simple relation Eq.(\ref{eq:magred}) is complicated by several effects. The lightcurve data are not available over the whole frequency range, but only in a small bandpass that has to be related to the whole bolometric luminosity. The adjustment to correct for this `bandpass mapping' is done by adding a correction term, whose value depends not only on the source's redshift but also on its spectrum \citep{Drell}. The main complication, however, seems to be the deduction of the object's intrinsic brightness, which cannot be measured directly, from its spectroscopic and/or photometric properties. While SNe Ia clearly do not always have the same brightness, but indeed have a significant dispersion of $0.5\mathrm{mag}$ in peak magnitude in most pessimistic estimations, one can establish a relation, however, between the decline rate $\Delta m_{15}$ (the total decline in brightness from peak to 15 days afterward) in the SN lightcurve shape and its absolute peak-brightness $M_{\mathrm{peak}}$:
\begin{equation}
\Delta m_{15} \propto M_{\mathrm{peak}}^{-1}
\label{deltam15}
\end{equation}
within an error margin of around $\sigma_{M_{\mathrm{intr}}}=0.15$ \citep{Kowalski}, which means distance estimates precise to $\approx 7\%$. Several techniques of standardization have been developed to implement this relation into SN data analysis, in particular the strech-factor method \citep{P99} and the template-fitting method \citep{Hamuy961,Hamuy962}. The possible introduction of systematic errors by the use of different fitting techniques have been exhaustively studied by \citet{Drell} and \citet{Guy}.\\

\noindent The use of Eq.(\ref{eq:magred}) would furthermore be complicated if the apparent luminosity $m$ was systematically varying with redshift due to various astrophysical effects.
\paragraph*{Intrinsic effects:} A first class of possible effects would affect the intrinsic properties of the SN, like the luminosity $M_{\mathrm{intr}}$ and the light-curve shape.  First indications of this kind of evolution of SN properties came from the measurement of shorter rise-times preceding maximum luminosity in SN light curves at higher redshift \citep{Riess99}. Various possible reasons for a redshift evolution of the intrinsic SN properties have been proposed. Because the younger universe is characterized by lower overall metallicity, different metallicity of SN progenitors at different cosmic epochs has been considered as possibly causing an overall evolutionary effect. Also the white dwarf's initial mass and different mass fractions of the composing elements at the moment of explosion, varying with cosmic time, are suspect of affecting the intrinsic SN properties. The effect of varying composition and metallicity of the SN progenitors has been studied numerically. The results of \citet{Hoeflich}, \citet{Lentz}, and \citet{Roepke} however disagree, in their predictions of $M_{\mathrm{intr}}$ and light-curve shape dependence on progenitor composition, but also use different types of explosion mechanisms (detonation/deflagration models).
Other effects might come from direct environmental phenomena like circumstellar gas clouds or magnetic fields. The progenitor's rotation is taken into account by three-dimensional numerical models, but the results rest unconclusive so far \citep{Roepke05}.
Furthermore, rather poorly understood dependencies of the overall energy output on the progenitor's age remain. \citet{Gallagher} point out that progenitor age might more likely be the source of variability in SN peak luminosities than is metallicity. A study of SN samples by the SN high-z team \citep{SNhighz} shows that SN models that require a large number fraction of prompt explosions (progenitor age $<$ 2 billion years) poorly reproduce the observed redshift distribution and are rejected at $> 95\%$ confidence. When looking at the young universe, at redshifts $z>1.5$, one clearly expects a younger mean progenitor sample, and a study of variation in SN magnitude with respect to the mean progenitor age in different host galaxies by \citet{Howell07} indeed suggests more luminous SNe at higher redshift.\\
\indent Another problem with the progenitor age arises: the two principal models of supernova Ia progenitor systems, \ie the single-degenerate and the double-degenerate progenitor system, differ in explosion delay time (and environmental conditions). The uncertainty in identifying the progenitor systems may mimick an evolutionary effect \citep{Riess06}.\\
\paragraph*{Subpopulations:} \citet{Hamuy961,Hamuy962} suggest that SN magnitude depends on the host galaxy (brighter SNe in early-type elliptic galaxies than in late-type spiral galaxies), and proposed a classification of SNe Ia into two subclasses. The study of \citet{Hatano} also indicates two subpopulations that may be identified as resulting from different explosion mechanisms, \ie plain detonation, deflagration, or delayed detonation models. Progenitor population drift leads to different SN subpopulations dominating at different cosmological epochs and may translate into an overall evolutionary effect \citep{Branch01}. The subpopulation bias has recently been studied by \citet{SarkarSubpopulations} and \citet{Linder09}, showing that subpopulations with different mean peak magnitudes may bias the parameter of dark energy equation of state.\\
\paragraph*{Extrinsic effects:} Another class of possible effects leading to an overall redshift evolution of apparent SN magnitude $m$ and light-curve shape consists of extrinsic effects, which do not come from the intrinsic SN properties being altered, but from astrophysical effects between the SN and the observer. Gravitational lensing effects have been considered to possibly mimick an SN magnitude evolution on cosmic time scales, but have been shown to be negligible \citep{SarkarLensing}. Also gray intergalactic dust has been proposed to mimick an evolutionary effect since \citet{Aguirre}, and has been shown to possibly bias the extraction of the cosmological parameters \citep{Corasaniti,Menard2009}. Also effects of the interstellar medium in the host galaxy may translate into evolutionary effects, the former systematically varying from early-epoch elliptic galaxies to late-epoch spiral galaxies.\\
\paragraph*{} Given this rather rudimentary understanding of SN magnitude evolution in redshift range say $0<z<1.7$, we find it reasonable to test the optimistic assumption that the calibration relations obtained for `local', low-redshift SNe be without modification valid for SNe at arbitrarily high redshift. We do not aim to extract evolutionary parameters or rule out models of SN magnitude evolution, but to study the systematics that are introduced when evolution is neglected in the analysis.\\
To this end we developed detection criteria and tested their performance when confronted with the wrong model assumption on the SN magnitude. In section \ref{sec:parameterizations} we introduce some phenomenological models to describe systematical shifts in SN magnitudes, discuss general aspects of SN magnitude evolution, and give limits on the model parameters obtained from current observational data in section \ref{sec:realdata}. In the main part, section \ref{sec:simulations}, we study the implications, an unaccounted for evolutionary effect on SN magnitudes would have on data analysis. Therein we make use of a simulated SN sample consisting of $\sim2000$ SNe up to redshift $z=1.7$. We present our approach to the detectability of such an effect, and discuss the possible introduction of nondetectable biases on the cosmological parameters. From merged detectability and bias results, we introduce the concept of the danger of the phenomenological models.  This paper ends with section \ref{sec:discussion}, where the results are summed up and discussed.

\section{\label{sec:parameterizations}Parameterizations}

The possible evolution of SN magnitude with redshift is allowed for by a simple modification of the magnitude-redshift relation, \ie in Eq.(\ref{eq:magred}) we allow for a magnitude altering $\Delta m\left(z\right)^{\mathrm{evo}}$,
\begin{equation}
m=M_{\mathrm{s}}+5\log{D_{\mathrm{L}}(z;\Omega_i,w_{\mathrm{x}})}+\Delta m(z)^{\mathrm{evo}},
\label{eq:magredevolution}
\end{equation}
supposed to originate in an overall time evolution of the SN peak luminosity. In the following we simply type $\Delta m^{\mathrm{evo}}$ instead of $\Delta m\left(z\right)^{\mathrm{evo}}$. Several phenomenological models for an SN magnitude evolution have been used in the literature. Because we know little about the physics of a magnitude variance with cosmic time, these parameterizations somehow are a `matter of taste' and are differently motivated according to the purpose of the respective analysis. \citet{Linder2006a} studies a $\Delta m^{\mathrm{evo}}\propto z/\left(1+z\right)$ model for reason that it is bounded at high redshift $z \rightarrow \infty$. \citet{Ferramacho} use a $\Delta m^{\mathrm{evo}} \propto \left(t\left(z\right)-t_0\right)$ in real data analysis, where $t$ is the cosmic time:
\begin{equation}
t(z;\Omega_i,w_i)=\frac{1}{H_0}\int\limits_z^{\infty}{\frac{1}{(1+\tilde{z})E(\tilde{z};\Omega_i,w_i)}\mathrm{d}\tilde{z}}.
\label{eq:cosmictime}
\end{equation}
In the $\Lambda CDM$ case, Eq.(\ref{eq:cosmictime}), however, is well approximated by a $\Delta m^{\mathrm{evo}}\sim\sqrt{z}$ model in the matter-dominated epoch $z>z_{\mathrm{eq}}$, where $z_{eq}$ is the redshift where $\Omega_{\mathrm{M}}\left(z\right)=\Omega_X\left(z\right)$. In the dark-energy dominated regime $z<z_{\mathrm{eq}}$ Eq.(\ref{eq:cosmictime}) is equivalent to $\Delta m^{\mathrm{evo}}\sim\log\left(1+z\right)$. This logarithmic model was first considered by \citet{Drell}, who motivated it as a parameterization of the intrinsic luminosity $\mathcal{L}$:
\begin{equation}
\mathcal{L}\rightarrow \mathcal{L}(1+z)^{-\beta_1},
\label{eq:drellparamL}
\end{equation}
where $\beta_1$ is a real number. This assumption on a possible redshift dependence of the intrinsic SN property trivially translates into the parameterization
\begin{equation}
\Delta m^{\mathrm{evo}} = 2.5\beta_1\log(1+z),
\label{eq:model1}
\end{equation}
which will be referred to as Model $1$ throughout this study. We in some respects prefer this consideration of a varying $\mathcal{L}$ instead of modifying $m$, for reason of the latter being an artificial construct.  Since we would expect a cosmological drift of SN magnitudes at very high redshift to come from astrophysical effects more or less \emph{directly} related to cosmological expansion, and assuming that $\mathcal{L}$ itself is more or less a direct function of these variables, we would expect the effect on $\mathcal{L}$ to be naturally related to the scale factor of the universe, $a\left(t\right)$, and thus to $\left(1+z\right)$. 
Being aware of the parameterization dependence of all analysis results, we complement our study by considering the general models
\begin{equation}
\Delta m^{\mathrm{evo}} = \beta_2 z^{\alpha}
\label{eq:models2}
\end{equation}
to study the contribution of powers in $z$ to our results. $\alpha$ and $\beta_2$ again are real numbers. For $\alpha \rightarrow 0$ in Eq.(\ref{eq:models2}) the evolutionary effect becomes highly degenerate with $M_{\mathrm{s}}$.
In particular, we use the three submodels $\alpha=0.5,\,1$, and $2$, which will be referred to as Model $2.1$, Model $2.2$, and Model $2.3$, respectively. These three power contributions are chosen in order to give more weight either to low-z or to high-z SNe in the sample. Model $2.1$ would model a magnitude evolution dominant at lower redshift $\left(z\la1\right)$, meaning a \emph{late-epoch model}. Model $2.3$ would give a magnitude evolution dominating at high redshift $\left(z\ga1\right)$ and will thus be considered as a \emph{early-epoch model}. We note that Model $2.2$ had already been the point of a qualitative discussion in the work of \citet{Riess}, and was studied by \citet{Nordin}. We also note that a Model $2$ with $\alpha\approx0.6$ describes the functional form of Eq.(\ref{eq:cosmictime}) correctly in the whole redshift range $0<z<1.7$, and at the same time is a good approximation to the $\propto z/\left(1+z\right)$ parameterization.

\section{\label{sec:realdata}Real data}

\subsection{\label{sec:constraints}Constraints on evolution}
We obtained current constraints on the $\beta$ parameters by a combined fit on the values of the cosmic microwave background (CMB) shift parameter $R=1.71\pm0.03$ \citep[see][]{Efstathiou,Wang}, the baryonic acoustic oscillation (BAO) parameter $A=0.469\pm0.017$ \citep[see][]{Eisenstein05}, and the 307 SN magnitudes as compiled by \citet{Kowalski}. Instead of a full CMB and BAO analysis, $R$ and $A$ are being used to save computing time, cf. section \ref{sec:simulations}. We impose flatness $\sum_i{\Omega_i=1}$ on all fits. To take a possibly time-varying value of the equation of state of the dark energy component into account, we used the Chevallier-Polarski-Linder \citep[CPL,][]{CP,L} parameterization $w_{\mathrm{x}}(z)=w_0+w_{\mathrm{a}}z/\left(1+z\right)$, where $w_{\mathrm{a}}=-2\frac{\mathrm{d}w}{\mathrm{d}\ln{a}}\big\vert_{z=1}$ depicts the overall time variation of $w_{\mathrm{x}}$. The fit parameters thus are $\{p_i\}=\{M_{\mathrm{s}},\Omega_{\mathrm{M}},w_0,w_{\mathrm{a}},\beta\}$. We neglected the radiation component $\Omega_R$ in the fit.\\

\begin{figure}
\begin{center}
\resizebox{\hsize}{!}{\includegraphics{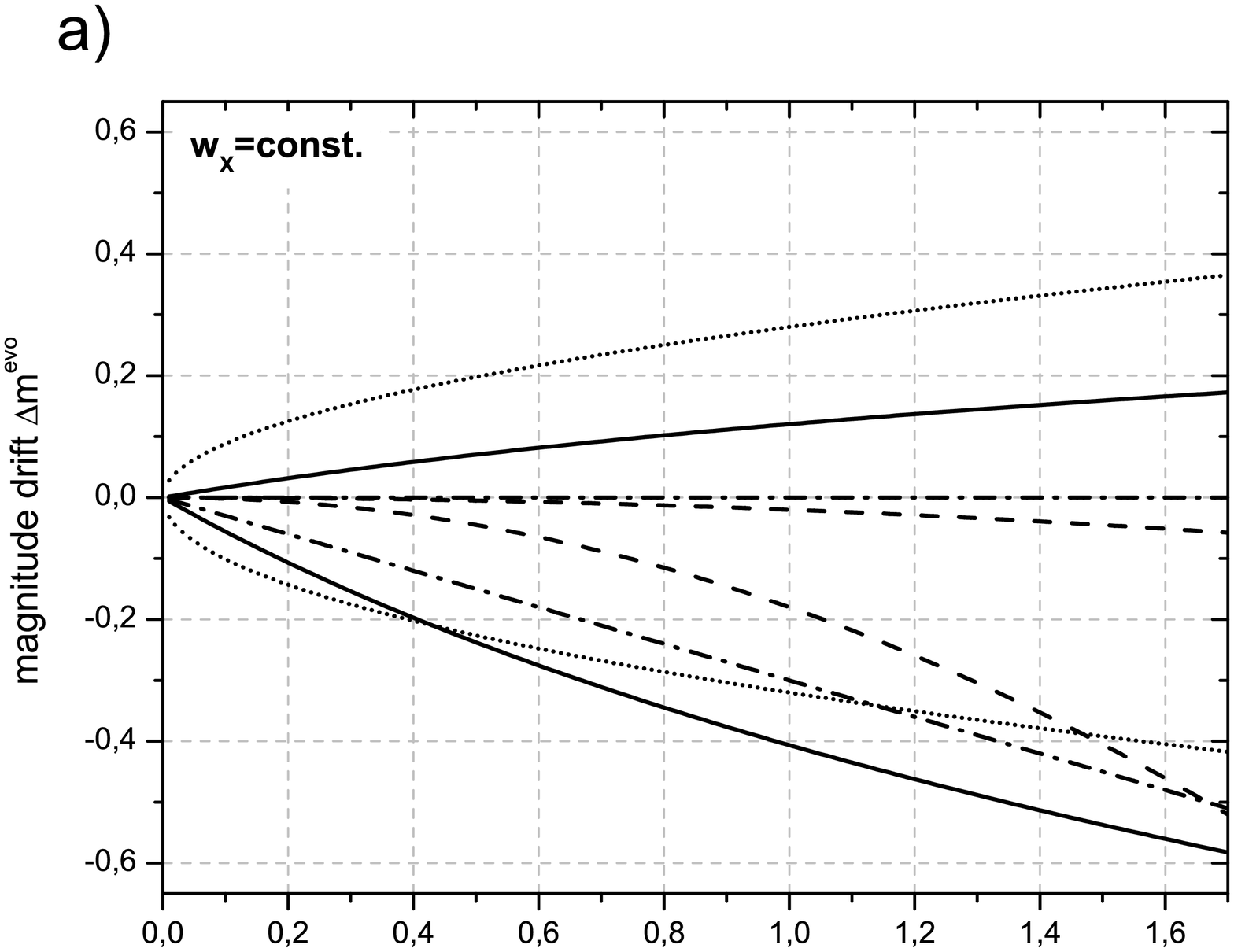}}
\resizebox{\hsize}{!}{\includegraphics{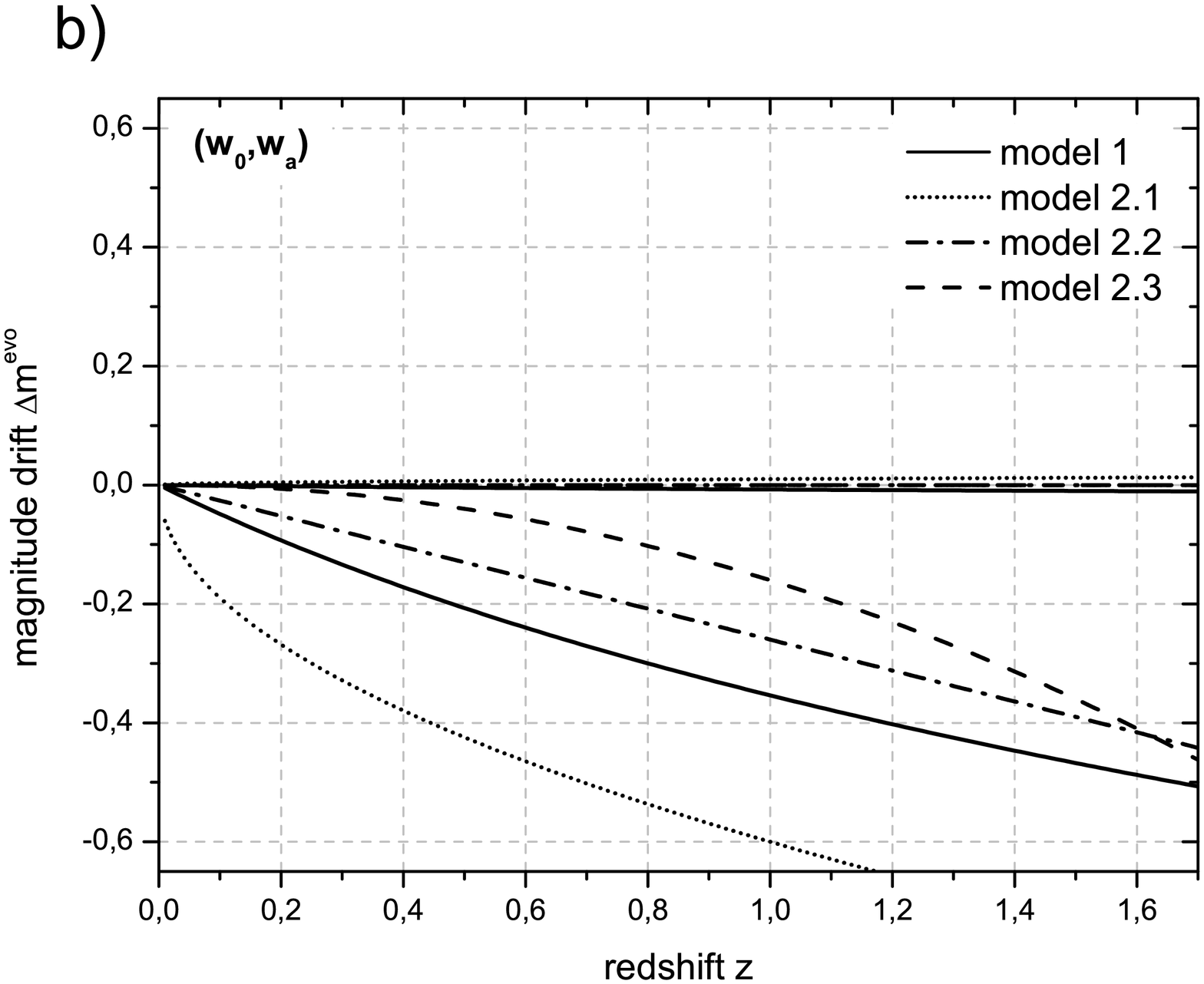}}
\caption{\label{fig:bestfitillustration} Illustration of the $1\sigma$ limits for models $1$ and $2$ obtained by an SNe+$R$+$A$ fit in a magnitude vs. redshift diagram. a) Dark energy equation of state $w_{\mathrm{x}}=\mathrm{const.}$ has been imposed in the fit. b) $w_{\mathrm{x}}=w_0+w_az/(1+z)$ is allowed to be dynamical.}
\end{center}
\end{figure}

\noindent The fit was performed with the cosmological analysis tool \upshape{GFIT}.\footnote{\upshape{GFIT} is a cosmological analysis tool developed by A.T. that allows studies of real or simulated data for various probes (SNe, CMB, BAO, weak lensing (WL), clusters, ...). Information on the code and its utilization can be obtained by contacting: tilquin@cppm.in2p3.fr.} The best fit is found by a minimization of the 
\begin{equation}
\chi^2=\left[\vec{m}-\vec{m}^{\mathrm{fit}}(z;p_i)\right]^T \vec{V}^{-1}\left[\vec{m}-\vec{m}^{\mathrm{fit}}(z;p_i)\right],
\label{eq:chi2}
\end{equation}
where $\vec{m} = (m_i,R,A)$ is the vector of the SN magnitudes $m_i=\{m_1,...,m_{307}\}$ and $R$ and $A$. The vector $\vec{m}^{\mathrm{fit}}\left(z;p_i\right)$ is the corresponding vector of values for the fitted parameters, and the covariance matrix $\mathbf{V}$ is the inverse of the error matrix whose elements are made up of the errors $\sigma_i$ on SN magnitudes and on $R$ and $A$: ${V_{ii}}^{-1}={\sigma_i}^2$. The errors on parameters $p_k$ are obtained by minimization over all other respective parameters $\{p_i\}_{i\neq k}$ for $p_k$'s in an environment around the best-fit value $p_k^{\;\mathrm{fit}}$. One sigma errors $\pm1\sigma_{p_k}$ then correspond to $\chi^2_{{\mathrm{min}}}+1$.\\
\begin{center}
\begin{table}[h!]
\caption{\label{tab:fitresults} Best-fit values for models $1$ and $2$.}
\begin{tabular*}{0.5\textwidth}{@{\extracolsep{\fill}}r|cccc|c}
\hline
\hline
\textbf{a)}\hspace{1.1cm} & $\Omega_{\mathrm{M}}$ & $\beta$ & \multicolumn{2}{c|}{$w_{\mathrm{x}}=\mathrm{const.}$} & $\chi^2/n$ \\
\hline
            i) & $0.27^{+0.02}_{-0.02}$ & $-$ & \multicolumn{2}{c|}{$-0.95^{+0.08}_{-0.08}$} & 0.72 \\
Model $\;\;\,1$& $0.24^{+0.05}_{-0.04}$ & $-0.22^{+0.38}_{-0.32}$ & \multicolumn{2}{c|}{$-1.18^{+0.34}_{-0.34}$} & 0.71 \\
Model $2.1$    & $0.29^{+0.05}_{-0.05}$ & $+0.08^{+0.20}_{-0.40}$ & \multicolumn{2}{c|}{$-0.86^{+0.24}_{-0.52}$} & 0.71 \\
Model $2.2$    & $0.24^{+0.04}_{-0.03}$ & $-0.15^{+0.15}_{-0.15}$ & \multicolumn{2}{c|}{$-1.16^{+0.22}_{-0.22}$} & 0.71 \\
Model $2.3$    & $0.26^{+0.02}_{-0.02}$ & $-0.10^{+0.08}_{-0.08}$ & \multicolumn{2}{c|}{$-1.00^{+0.12}_{-0.12}$} & 0.71 \\
\hline
& & & & & \\
\multicolumn{6}{c}{}\\
\textbf{b)}\hspace{1.1cm} & $\Omega_{\mathrm{M}}$ & $\beta$ & $w_0$ & $w_{\mathrm{a}}$ & $\chi^2/n$ \\
\hline
            ii)& $0.28^{+0.02}_{-0.02}$ & $-$ & $-1.14^{+0.16}_{-0.08}$ & $0.88^{+0.49}_{-0.49}$ & 0.71 \\
Model $\;\;\,1$& $0.24^{+0.04}_{-0.04}$ & $-0.24^{+0.23}_{-0.23}$ & $-1.48^{+0.38}_{-0.31}$ & $1.39^{+0.40}_{-0.60}$ & 0.71 \\
Model $2.1$    & $0.23^{+0.05}_{-0.05}$ & $-0.30^{+0.31}_{-0.30}$ & $-1.65^{+0.55}_{-0.47}$ & $1.72^{+0.70}_{-1.03}$ & 0.71 \\
Model $2.2$    & $0.25^{+0.03}_{-0.03}$ & $-0.13^{+0.13}_{-0.13}$ & $-1.36^{+0.24}_{-0.24}$ & $1.19^{+0.51}_{-0.91}$ & 0.71 \\
Model $2.3$    & $0.26^{+0.02}_{-0.02}$ & $-0.08^{+0.08}_{-0.08}$ & $-1.21^{+0.18}_{-0.18}$ & $0.91^{+0.48}_{-1.09}$ & 0.71 \\
\hline
\end{tabular*}
\end{table}
\end{center}
The results for all four models are assembled in Table \ref{tab:fitresults}, where we show the results for the case $w_{\mathrm{x}}=\mathrm{const.}$, and the dynamical case $w_{\mathrm{x}}=w_0+w_{\mathrm{a}}z/\left(1+z\right)$, case a) and b), respectively. 
The table also shows the results when $\beta=0$ was imposed on the fit, cases i) and ii), respectively. These agree with the results of \citet{Kowalski}, \citet{Komatsu}, and \citet{Tilquin}. In the following we call the model with $\{\Omega_{\mathrm{M}},w_0,w_{\mathrm{a}}\}=\{0.28,-1,0\}$ the $\mathrm{\Lambda CDM}$ model.\\
Concerning the $\chi^2$ per degrees of freedom $n$ of the fit, all models of Table \ref{tab:fitresults} perform equally well, \ie there is no significant tendency toward considerably higher or smaller $\chi^2$ for any of the models.\\
When we impose $w_{\mathrm{x}}=\mathrm{const.}$, we see a surprisingly high best-fit value of $\beta_1=-0.22^{+0.38}_{-0.32}$ for Model $1$, which corresponds to a magnitude drift ${\Delta m^{\mathrm{evo}}}_{\vert1.7}=-0.24^{+0.41}_{-0.35}$ at redshift $z=1.7$. The $\Omega_{\mathrm{M}}$ best-fit value is decreased by $0.03$ compared to the standard $\beta=0$ fit, which gives $\Omega_{\mathrm{M}}=0.27$. All $\{p_i\}$, however, are consistent with $\mathrm{\Lambda CDM}$ at the $1\sigma$ level. Model $2.2$'s evolutionary parameter has a lower negative best-fit value and is consistent with non-evolutionary SN magnitudes at the $1\sigma$-level, while $w_{\mathrm{x}}$ in this model is closer to the $\mathrm{\Lambda CDM}$ value $-1$, and errors on the parameters $\beta$ and $w_{\mathrm{x}}$ are smaller. Model $2.3$ again shows smaller errors on the parameters $\beta$ and $w_{\mathrm{x}}$, and the $\Omega_{\mathrm{M}}$ and $w_{\mathrm{x}}$ parameters approaching the $\mathrm{\Lambda CDM}$ values $0.28$ and $-1$, respectively. Its evolutionary parameter $\beta$, however, shows up to be inconsistent with non-evolving SN magnitudes at $1\sigma$. Model $2.1$ is a special case in this table, since it is the only one with a positive best-fit value for $\beta$. But it is consistent with no evolution at all at $1\sigma$. We find that at $\alpha\approx0.6$ the sign of the best-fit $\beta$ turns around, and that we fit $\beta\grole0$ for all $\alpha\leogr0.6$. Due to correlation between $\beta$ and $w_{\mathrm{x}}$, the best fit value of $w_{\mathrm{x}}$ also changes qualitatively with decreasing $\alpha$, and we fit $w_{\mathrm{x}}\grole-1$ for all $\alpha\leogr0.6$.\\
\indent When we allow the dark energy equation of state to vary with time, \ie if we include $w_{\mathrm{a}}$ in the fit, the picture changes considerably. In Model $1$ the $w_0$ best-fit value drops significantly below $-1$. Equally the $w_{\mathrm{a}}$ value is high, and the fit becomes inconsistent with a cosmological constant $\{w_0,w_{\mathrm{a}}\}=\{-1,0\}$ at the $1\sigma$-level. For Model 1 the $w_a$ value even is inconsistent with its cosmological constant value at $2\sigma$. For models $2$ we find the tendency that higher powers in redshift results in lower $|\beta|$ best-fit values, and $\Omega_{\mathrm{M}}$ best fit-value increasing towards $\Omega_{\mathrm{M}}=0.28$. $w_0$ equally moves towards its $\mathrm{\Lambda CDM}$ value $-1$, but remains inconsistent with it at the $1\sigma$ level. The CPL parameter $w_{\mathrm{a}}$ is inconsistent with $w_{\mathrm{a}}=0$ also in models $2.1$ and $2.2$, but becomes consistent with $w_{\mathrm{a}}=0$ in Model $2.3$.
We note for Model $2.1$ that the introduction of the $w_{\mathrm{a}}$ parameter pushed the $\alpha=0.6$ limit, where the best-fit value of $\beta$ became positive and $w_0>-1$ in the $w_{\mathrm{x}}=\mathrm{const.}$ case, to lower redshift powers $\alpha\approx0.1$ due to the strong degeneracy between $\beta$ and $w_{\mathrm{a}}$. All models including Model $2.1$ have $\beta<0$ best-fit values.\\
\noindent Figure \ref{fig:bestfitillustration} illustrates the best-fit values of Table \ref{tab:fitresults} in a magnitude redshift diagram; \ie, we plot the magnitude drifts $\Delta m^{\mathrm{evo}}$ obtained from the best-fit $\beta$ values via Eqs. (\ref{eq:model1}) and (\ref{eq:models2}) over redshift. The figure highlights that all $\Delta m^{\mathrm{evo}}>0$ are excluded at $1\sigma$ when $w_{\mathrm{x}}\neq \mathrm{const.}$ in the fit. The data are consistent with $\Delta m^{\mathrm{evo}}=0$ in all the fits, with two exceptions. Model $2.3$ exludes the hypothesis of nonevolving magnitudes at $1\sigma$ in the case where a constant dark energy equation of state $w_{\mathrm{x}}=\mathrm{const.}$ is assumed. And Model $1$ does the same when $w_{\mathrm{a}}$ is included in the fit.\\

\noindent Table \ref{tab:allcontours} shows all contours we obtain by a Model $1$ fit, with the four different fit-parameter sets i), ii), a) and b), respectively, cf. Table \ref{tab:fitresults}. Contours in parameter spaces $\left(p_j,p_k\right)$ are obtained by a $\chi^2$-minimization over all other parameters $\{p_i\}_{i\neq j,k}$, and $\{68\%,95\%\}$ confidence level contours correspond to contours $\chi^2_{\mathrm{min}}+\{2.3,5.99\}$, respectively. The parameter pairs $p_j \times p_k$ in the headings of the columns depict the $\mathrm{abscissae}\, \times\, \mathrm{ordinates}$ of the contour plots. 
We see that introducing the evolutionary parameter $\beta_1$ in the fit results in a decrease of the best fit $\Omega_{\mathrm{M}}$ value, which as we saw above is valid for all models whenever $\beta^{\mathrm{fit}}<0$. Concerning the present value of dark energy's equation of state $w_0$, including the $\beta_1$ parameter in the fit has the same effect as including $w_a$, \ie it moves the $w_0$ best-fit value beyond the phantom barrier $w_{\mathrm{x}}=-1$. Whereas including \emph{either} $w_{\mathrm{a}}$ \emph{or} $\beta_1$ keeps the fit result consistent with $w_0=-1$, including both of them at the same time makes $w_0$ inconsistent with $w_0=-1$. This effect persists in all models, but is the strongest in the late-epoch model $2.1$ and the logarithmic model $1$.
\begin{center}
\begin{sidewaystable*}[h!]
\vspace{120ex}
\caption{\label{tab:allcontours} All $68\%$, and $95\%$ confidence level contours for Model $1$. Full lines represent the $\Lambda CDM$ values $\{w_0,w_a,\beta_1\}=\{-1,0,0\}$ and the $(w_0+w_{\mathrm{a}})=0$ constraint.}
\begin{tabular}{rcccccc}
\hline
\hline
\multicolumn{1}{c|}{} &  \multicolumn{1}{c|}{\begin{small}$ \Omega_{\mathrm{M}}\,\times\,w_0 $\end{small}} & \multicolumn{1}{c|}{\begin{small}$ \Omega_{\mathrm{M}}\,\times\,w_a $\end{small}} & \multicolumn{1}{c|}{\begin{small}$w_0\,\times\,w_a$\end{small}} & \multicolumn{1}{c|}{\begin{small}$ w_0\,\times\,\beta_1 $\end{small}}  & \multicolumn{1}{c|}{\begin{small}$\Omega_{\mathrm{M}}\,\times\,\beta_1$\end{small}} & \multicolumn{1}{c}{\begin{small}$ w_a\,\times\,\beta_1 $\end{small}} \\
\hline
\multicolumn{1}{c|}{} & & & & & & \\
\multicolumn{1}{c|}{\textbf{i)}}          &  \includegraphics[width=3.5cm]{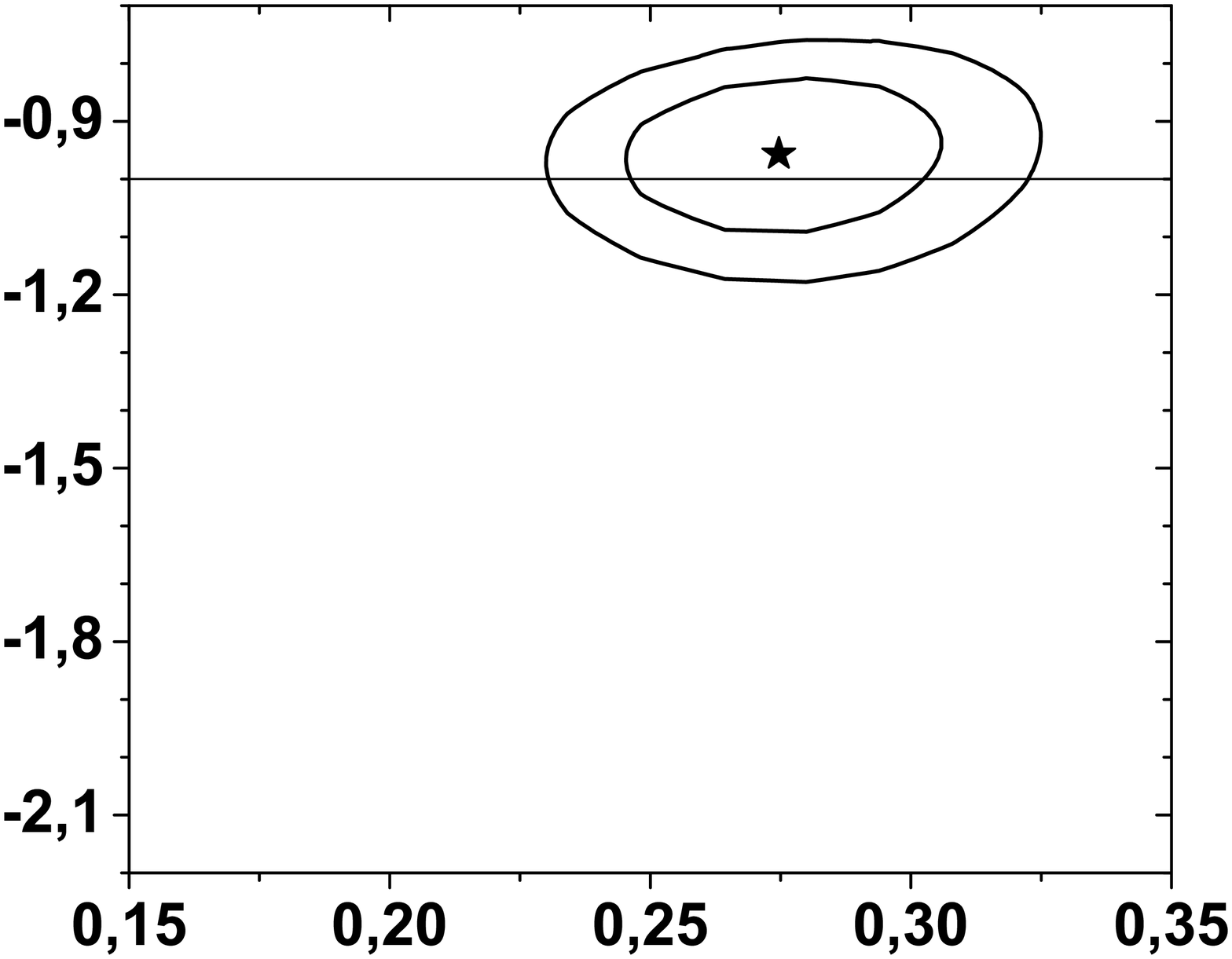} & & & & &  \\[1ex]
\multicolumn{1}{c|}{\textbf{ii)}}          &  \includegraphics[width=3.5cm]{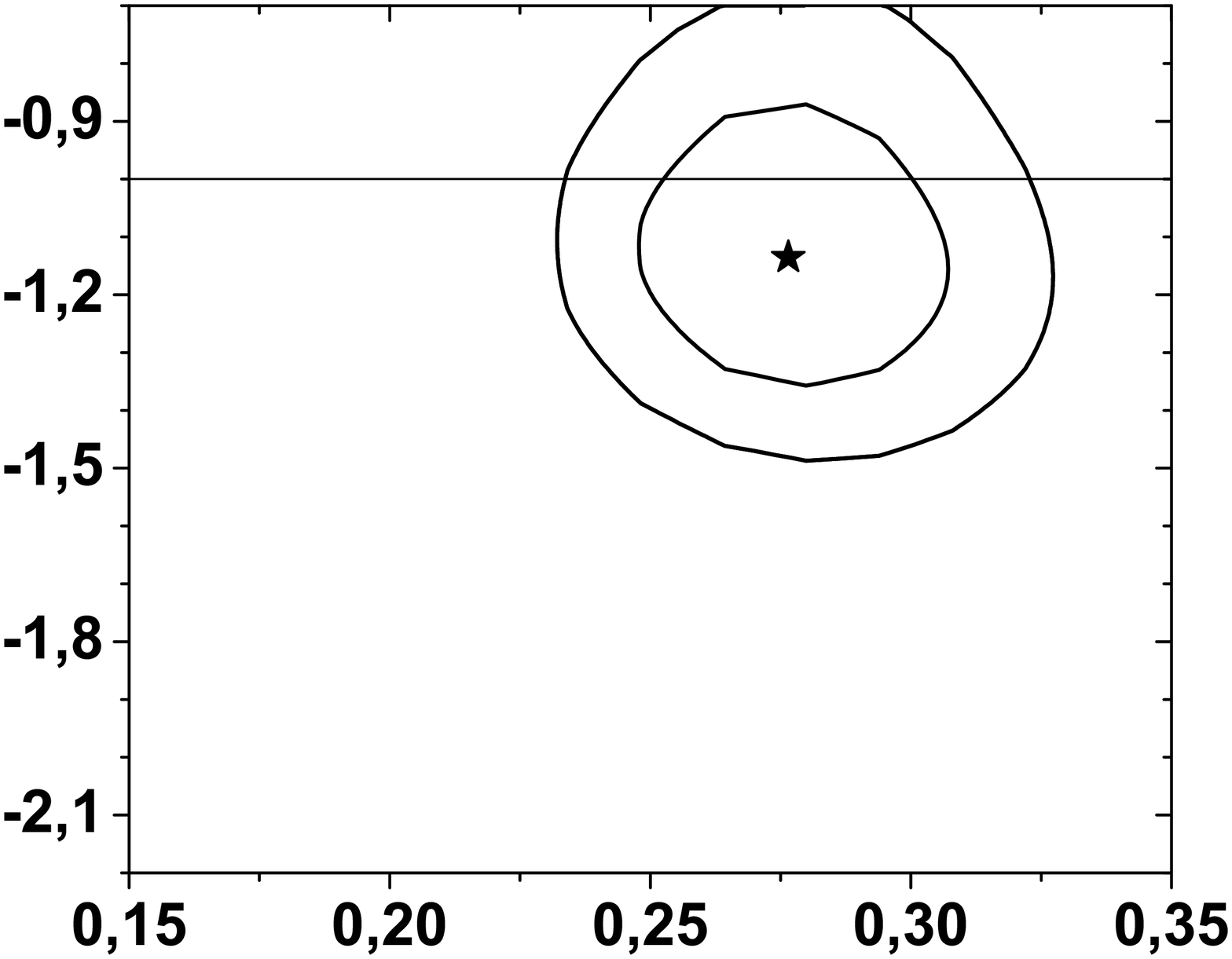} &  \includegraphics[width=3.5cm]{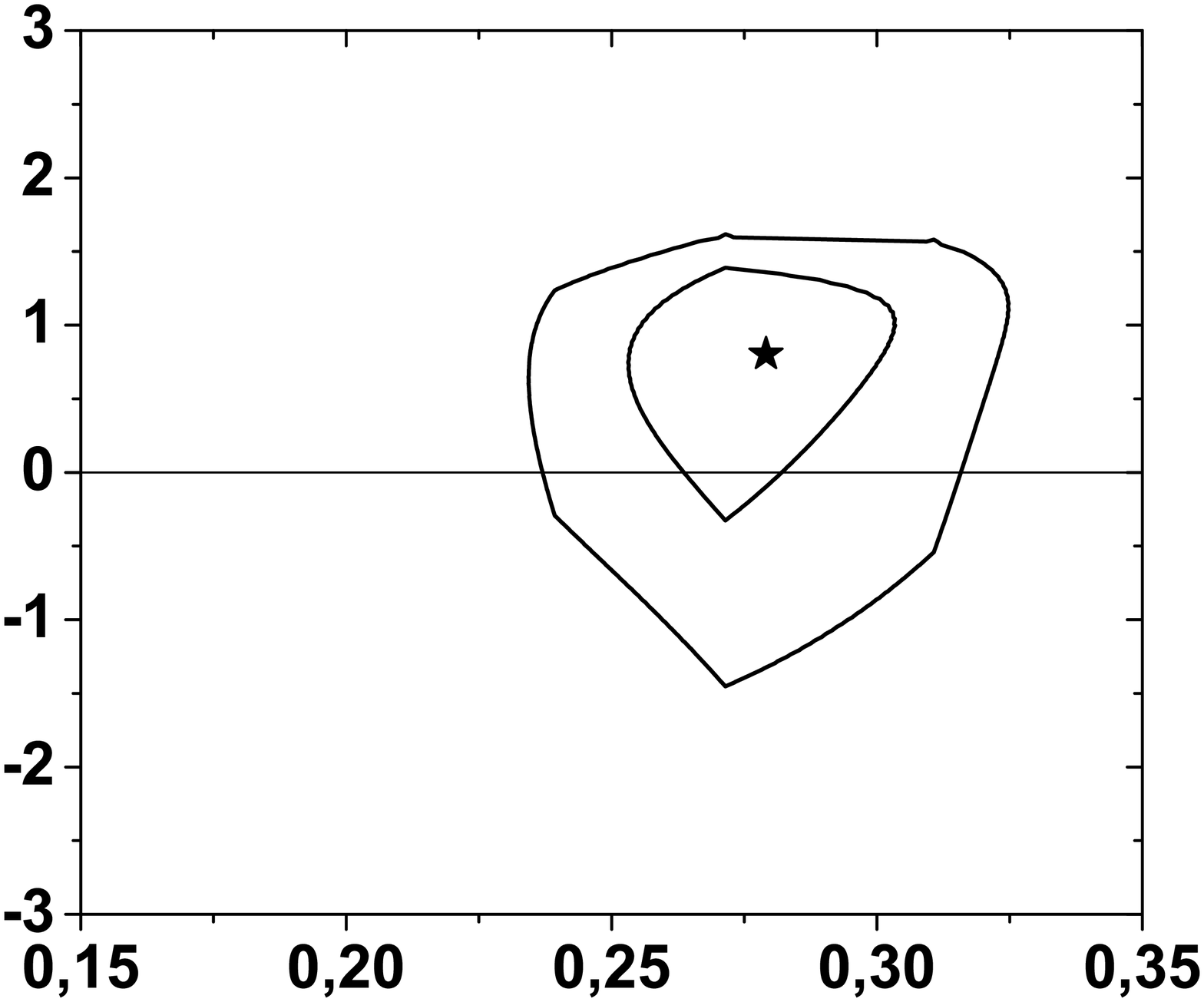} & \includegraphics[width=3.5cm]{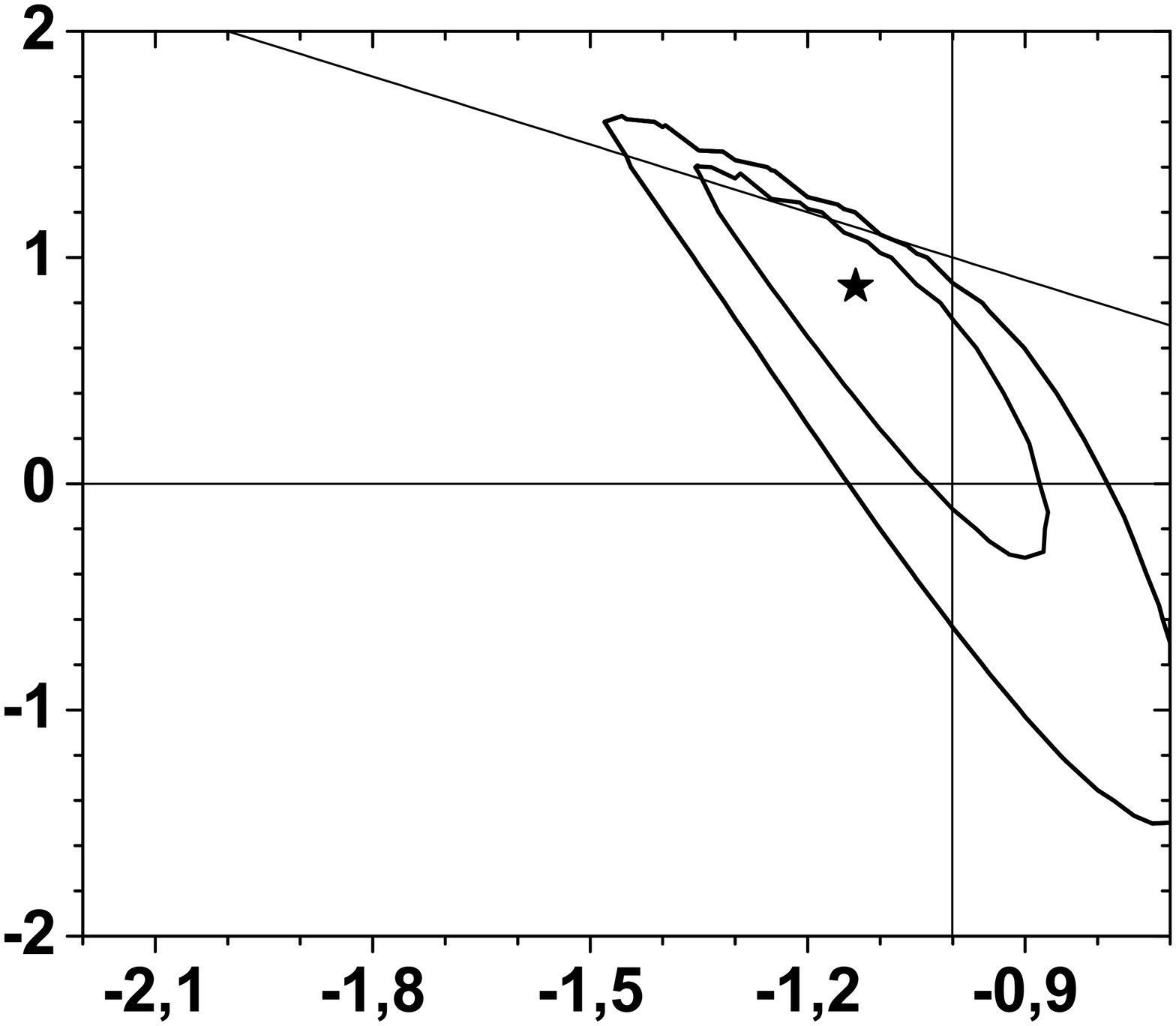} & & &  \\[1ex]
\multicolumn{1}{c|}{\textbf{a)}}          & \includegraphics[width=3.5cm]{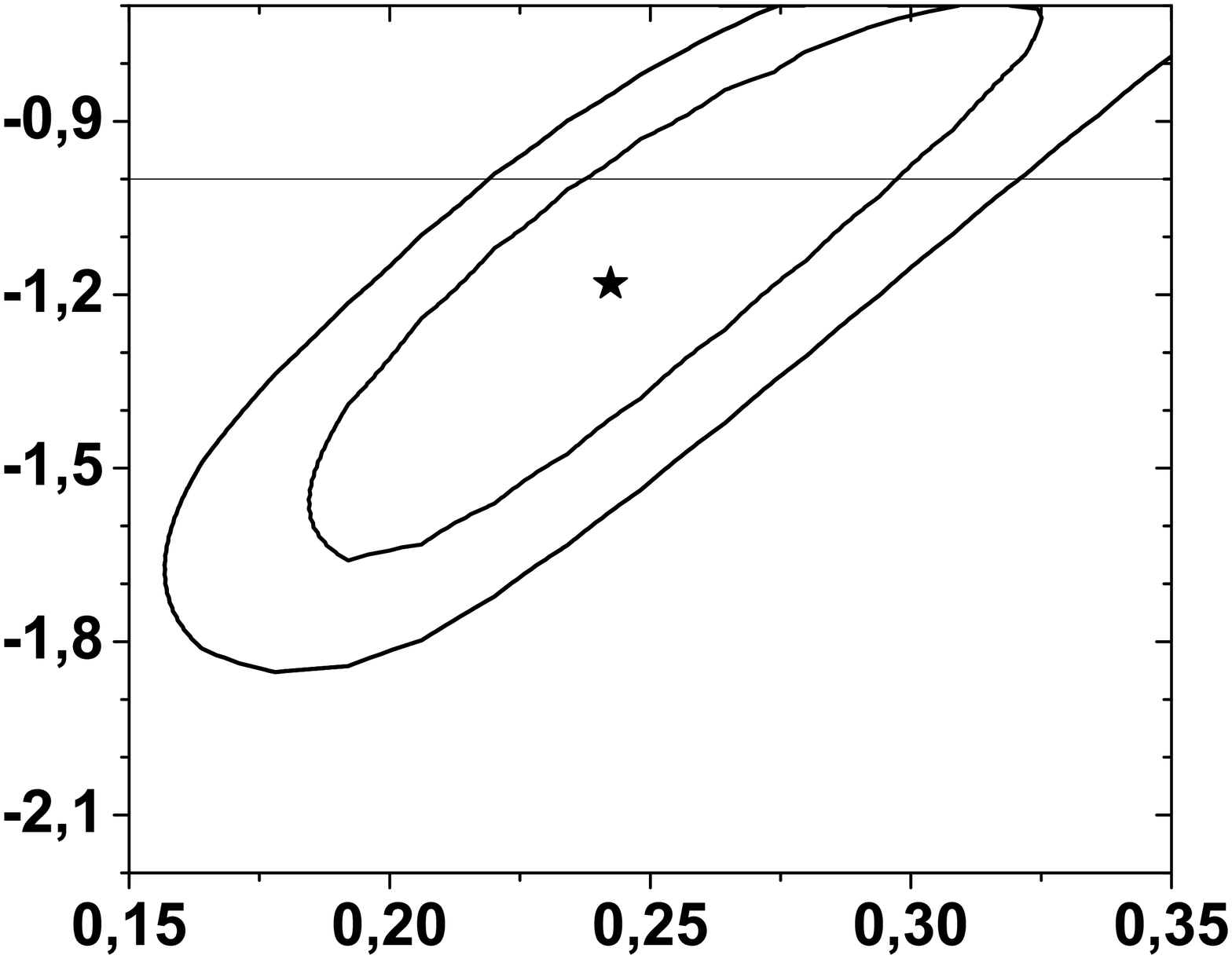} & &  & \includegraphics[width=3.5cm]{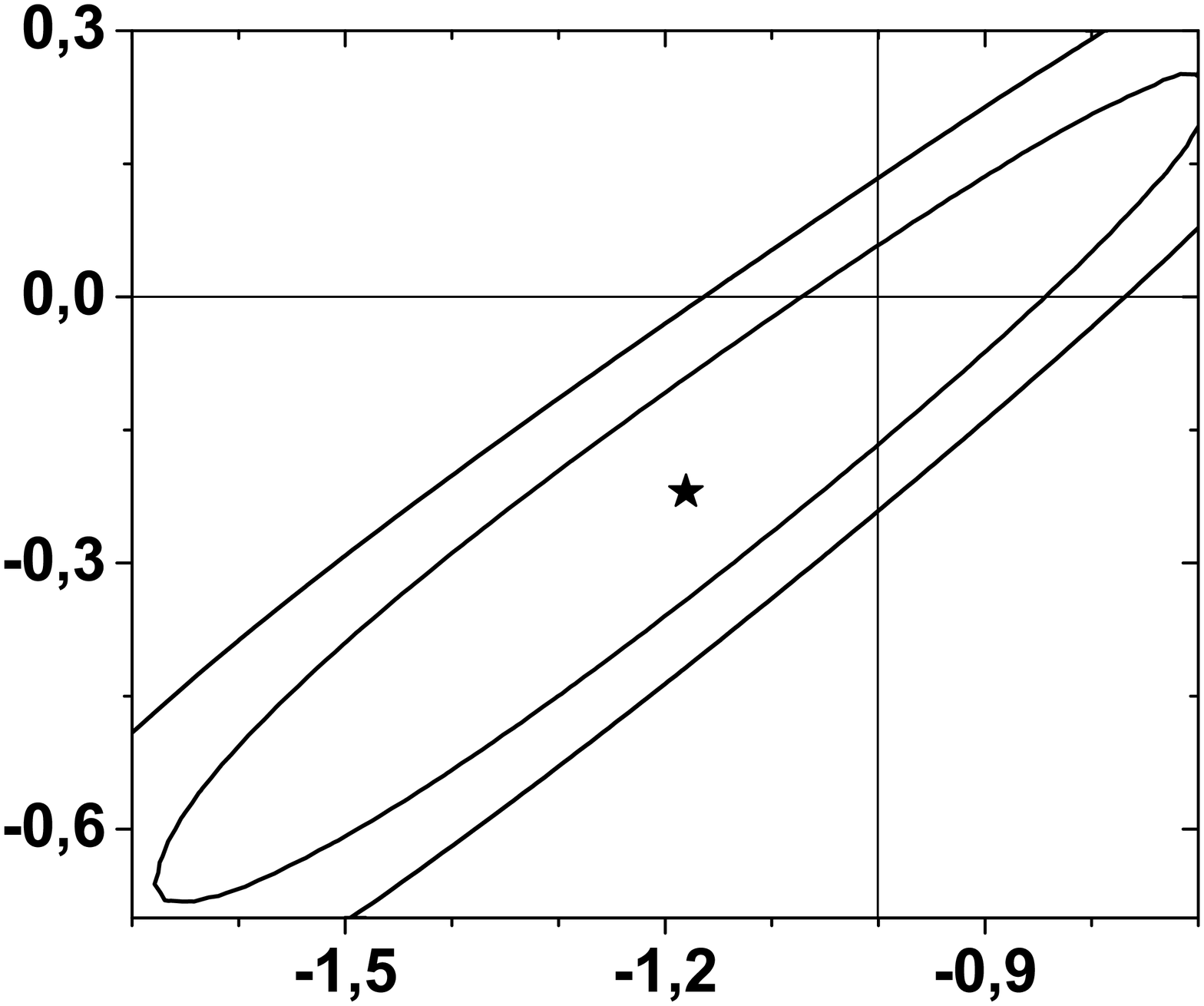} & \includegraphics[width=3.5cm]{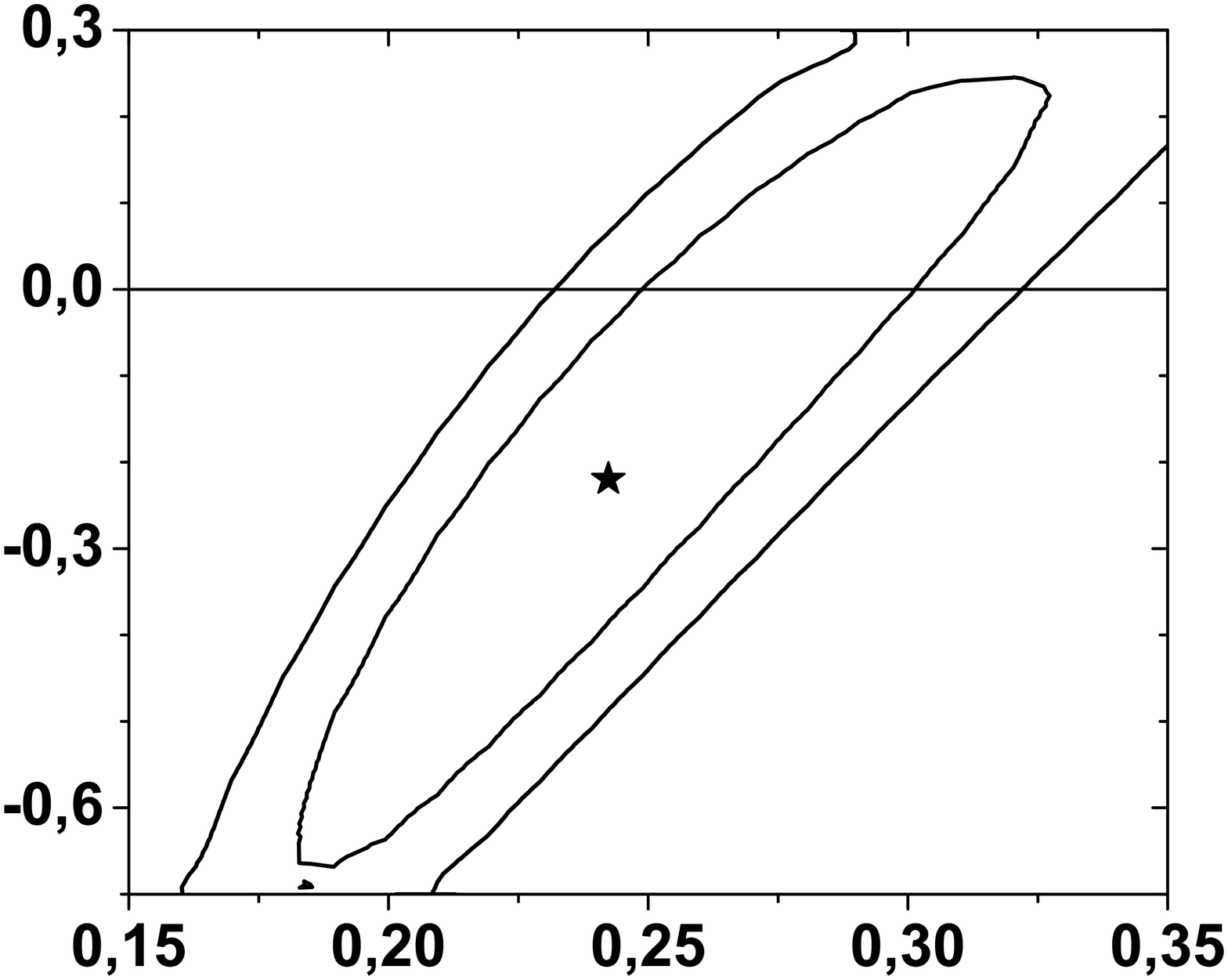} &  \\[1ex]
\multicolumn{1}{c|}{\textbf{b)}}          & \includegraphics[width=3.5cm]{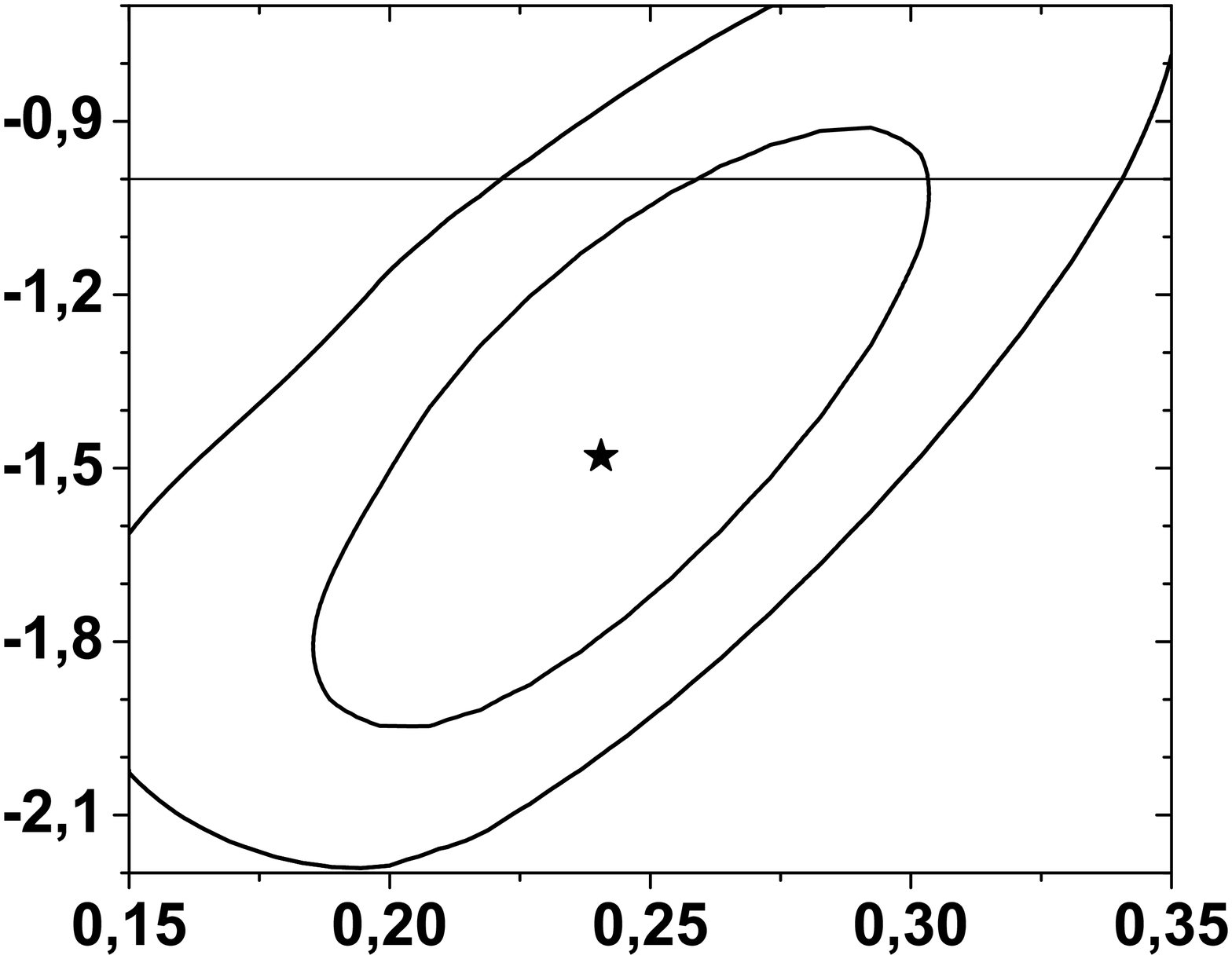} & \includegraphics[width=3.5cm]{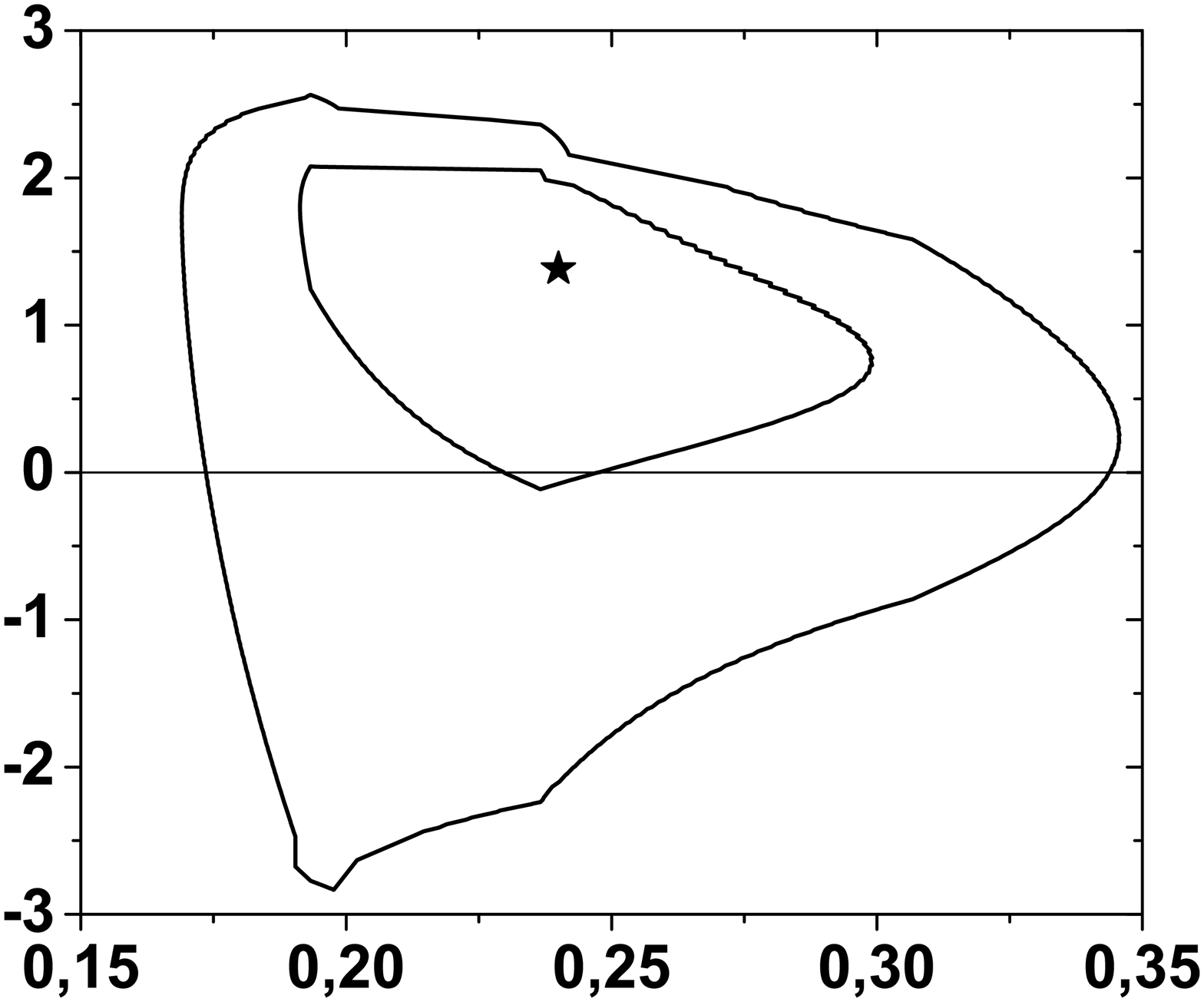} & \includegraphics[width=3.5cm]{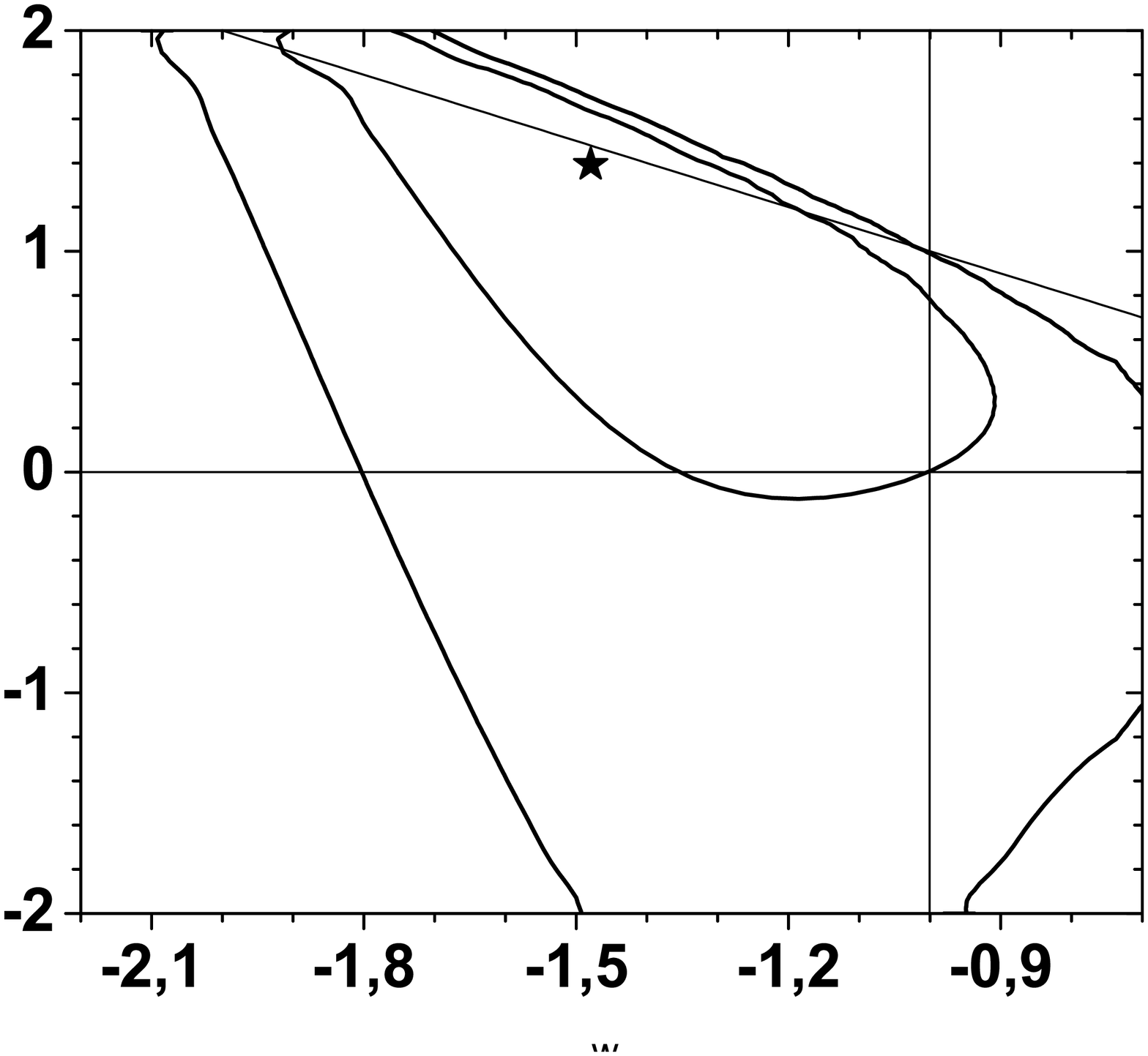} & \includegraphics[width=3.5cm]{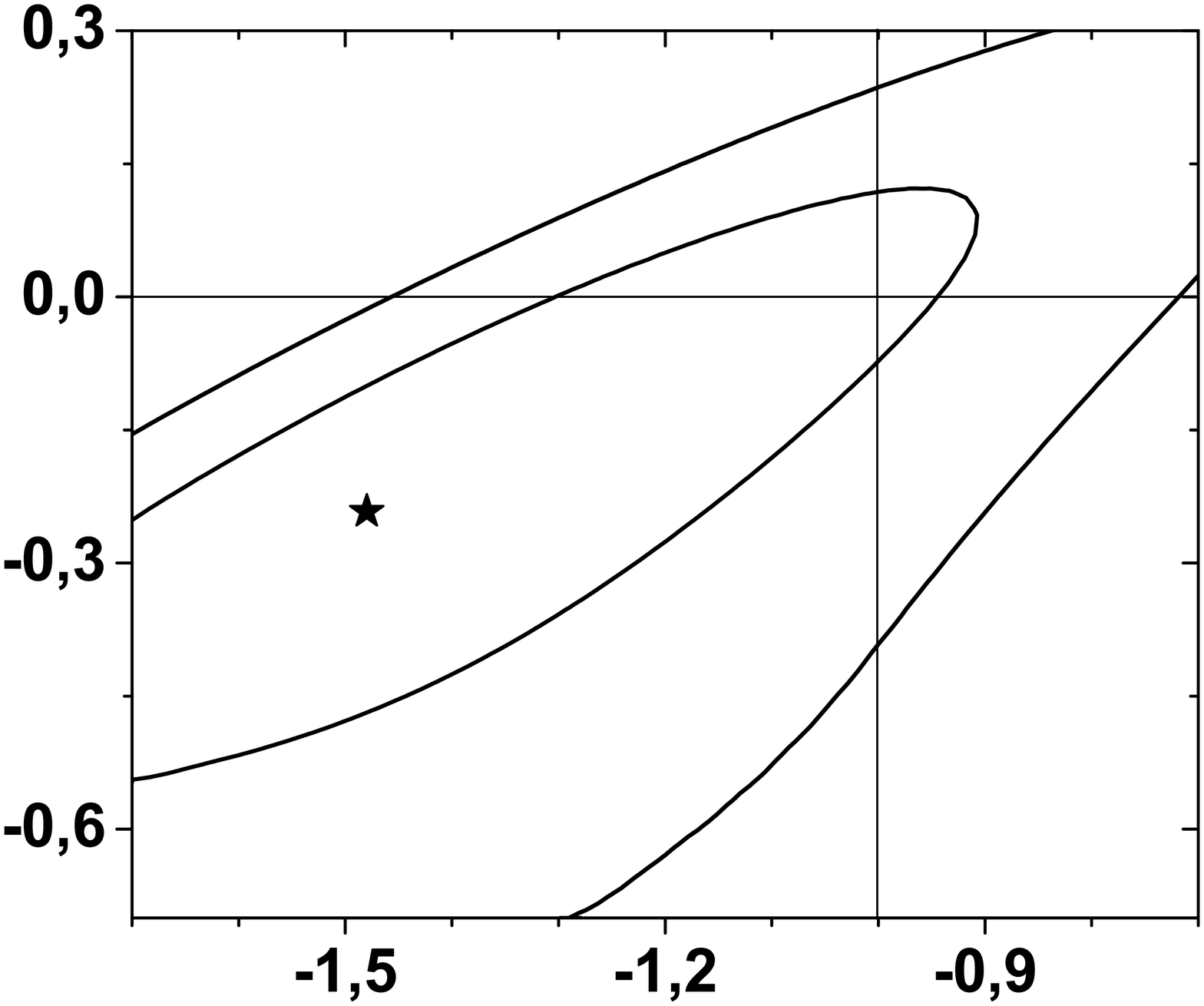} & \includegraphics[width=3.5cm]{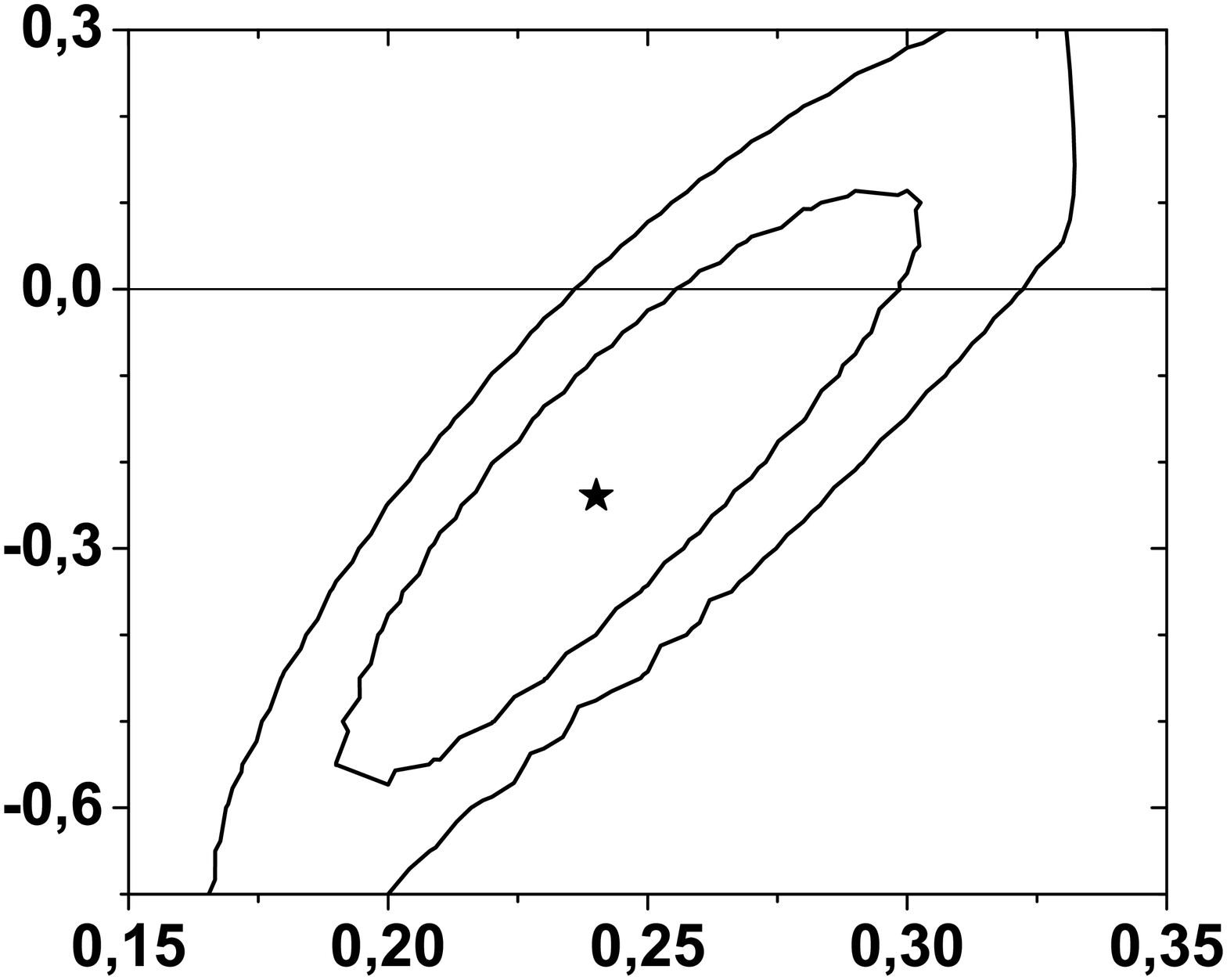} &  \includegraphics[width=3.5cm]{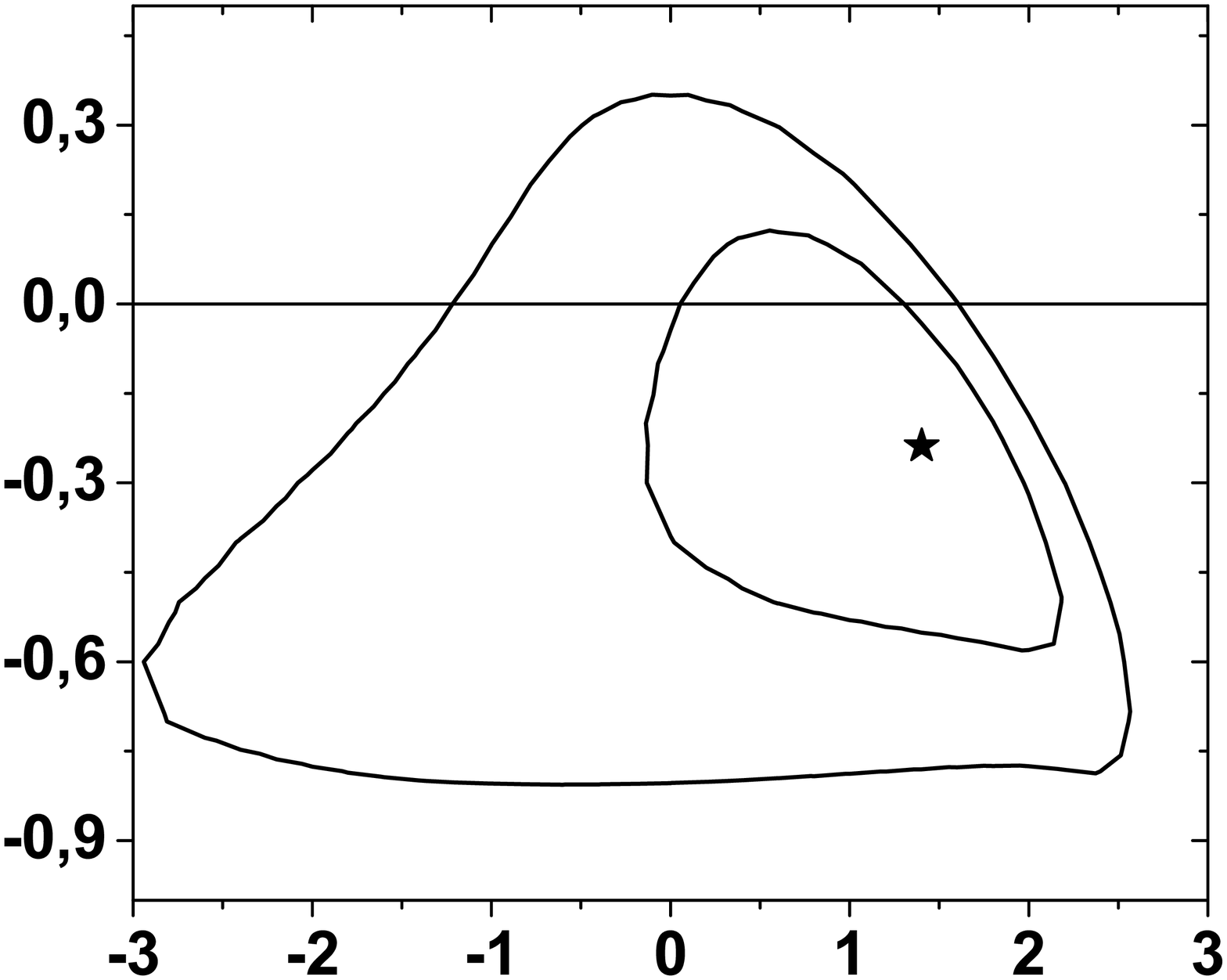} \\[3ex]
\multicolumn{1}{c|}{}& & & & & & \\
\hline
\end{tabular}
\end{sidewaystable*}
\end{center}

\subsection{\label{sec:HubbleDiagramm}Hubble diagram}

\begin{figure}[h!]
\begin{center}
\resizebox{\hsize}{!}{\includegraphics{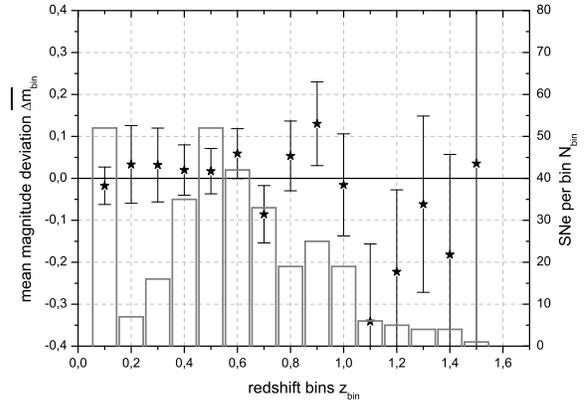}}
\caption{\label{fig:unionbinned} The relative magnitude deviation $\overline{\Delta m}_{\mathrm{bin}}$, Eq.(\ref{eq:deltambin}), of the SN sample from the magnitudes we obtain from the case ii) best fit.}
\end{center}
\end{figure}

The general tendency to prefer a negative $\beta$ (negative magnitude drift $\Delta m^{\mathrm{evo}}$) and thus brighter SNe at higher redshift can also be seen from the Hubble diagram of the SNe sample. We bin the sample in redshift bins of width $0.1$. Figure \ref{fig:unionbinned} shows the relative magnitude deviation 
\begin{equation}
\overline{\Delta m}_{\mathrm{bin}}=\frac{1}{N_{\mathrm{bin}}}\sum\limits_{i\in {\mathrm{bin}}}(m_i-m_i^{\;\mathrm{fit}})
\label{eq:deltambin}
\end{equation}
of the SNe sample from the best-fit magnitudes of case ii). Here, $N_{\mathrm{bin}}$ depicts the number of SNe in the bin. Error bars
\begin{equation}
\left(\frac{1}{\sigma_{\mathrm{m}_{\mathrm{bin}}}}\right)^2=\frac{1}{N_{\mathrm{bin}}}\sum\limits_{i\in {\mathrm{bin}}}\left(\frac{1}{\sigma_{\mathrm{m}_i}}\right)^2
\end{equation}
come from the errors $\sigma_{\mathrm{m}_i}$ on magnitudes as reported by \cite{Kowalski}. We included in gray the distribution of SNe in the plot, the right ordinate showing the number $N_\mathrm{{bin}}$ of SNe in the bin. High-redshift supernovae, \ie SNe at $z \ga 1$, seem to have a tendency to appear more luminous than the best fit. 
A linear least residual squares fit on this binned magnitude deviation gives a negative slope of
\begin{equation}
\frac{\Delta m^{\mathrm{evo}}}{z}=-0.04\pm0.06
\end{equation}
magnitude per redshift, which corresponds to a magnitude drift ${\Delta m^{\mathrm{evo}}}_{\vert1.7}=-0.07\pm0.10$ at redshift $z=1.7$. Data thus seem to prefer more luminous SNe at higher redshift, but are consistent with nonevolving SN magnitudes at $1\sigma$. We do, however, see clearly from Fig. \ref{fig:unionbinned} that with the current data the statistics is poor at high redshift, with only $18$ SNe at $z>1$. More high-redshift SN magnitudes are indispensable to check for magnitude evolution, as frequently pointed out since \citet{Riess}, \citep[\eg][]{Riess06}.\\

From sects. \ref{sec:constraints} and \ref{sec:HubbleDiagramm} we draw the conclusion that, while preferring negative magnitude drifts ${\Delta m^{\mathrm{evo}}}_{\vert1.7}<0$, data show no definite indication of SN magnitude evolution, in agreement with \citet{Kowalski}, \citet{Ferramacho}, \citet{Linder09}, and the results reported by \citet{SNLS} and \citet{Sullivan09}.

\section{\label{sec:simulations}Simulations}
We now study the impact, an unaccounted for SN magnitude evolution would have on the values of the extracted cosmological parameters in the framework of a combined analysis. Our database is a set of $2000$ simulated SN magnitudes in redshift range $0<z<1.7$, from which $\sim1000$ at $z\ge1$. The sample corresponds to what could be expected from a satellite mission with good control of systematics \citep{Kim}. We added $300$ SNe in the nearby range $z<0.1$ as expected by nearby SN Ia surveys like the Nearby Supernova Factory \citep{SNFact}. The sample was binned into redshift bins of width $0.1$. The intrinsic error on each SN magnitude is assumed to be $\sigma_{\mathrm{intr}}=0.15$, and systematic errors $\sigma_{\mathrm{sys}}=0.02$ on magnitudes are included in the analysis. This systematic error enters the covariance matrix by
${V_{ii}}^{-1}={\sigma_i}^2+(\sigma_i)_{\mathrm{sys}}^2$ and ${V_{ij}}^{-1}=r_{ij}(\sigma_i)_{\mathrm{sys}}(\sigma_j)_{\mathrm{sys}}$, where $r_{ij}$ is the correlation coefficient between bins, which however is set to $r_{ij}=0$. In the simulation of the SN magnitudes we allowed for a redshift evolution according to Eqs. (\ref{eq:model1}) and (\ref{eq:models2}). Scan range in $\beta$ is $-0.5<\beta<0.5$ for Model $1$ and Model $2.1$, $-0.4<\beta<0.4$ for Model $2.2$, $-0.3<\beta<0.3$ for Model $2.3$, and the other respective fiducial cosmological parameters are fixed to $\{\Omega_{\mathrm{M}},w_0,w{\mathrm{a}}\}^{\mathrm{Fid}}=\{0.30,-1,0\}$.\footnote{In cases where a confusion is possible, we added a superscript `Fid' to fiducial parameters and a superscript `fit' to fitted parameters.} We checked the stability of our results to variations in these fiducial parameters and found no significant variance from our general conclusions. Fit parameters are the standard parameters $\{p_i\}^{\mathrm{fit}}=\{M_{\mathrm{s}},\Omega_{\mathrm{M}},w_0,w_{\mathrm{a}}\}$. The magnitude evolution inherent in the simulated data was thus neglected in the fitting procedure; \ie, we imposed $\beta=0$ on the fit. To economize computing time we did not perform a combination with full BAO and CMB data, but simulated the CMB shift parameter $R$ and the BAO reduced parameter $A$. We find this simplification justified, because our task is not to extract real cosmological parameter values but to study the systematics introduced by a possibly wrong assumption on one of the probes. We assumed an error of $\pm0.01$ on $R$ (which is the estimate for future PLANCK data) and an error of $\pm0.005$ on $A$, to represent an experimental setup available in the near future.\\
The numerical tool for our parameter study with simulated data is \upshape{KOSMOSHOW},\footnote{\upshape{KOSMOSHOW} is a cosmological analysis tool developed by A.T., and is available on http://marwww.in2p3.fr/renoir/Kosmo-Pheno.php3.} which implements the $\chi^2$-minimization procedure as described in section \ref{sec:realdata}. The errors on the cosmological parameters $\{p_i\}$ are estimated at the minimum by using the first-order error propagation technique: $\vec{U} = \vec{J}.\vec{V}.\vec{J}^T$, where $\mathbf{U}$ is the error matrix on the cosmological parameters, and $\mathbf{J}$ the Jacobian of the transformation. The fit of the simulated data gives the central values $\{p_i\}$ and errors $\{\sigma_{p_i}\}$ for the parameters, along with their correlations $\rho_{ab}$, which in general are strong. We neglect the radiation component $\Omega_{\mathrm{R}}$ and continue assuming spatial flatness in the following, $\sum_i{\Omega_i = 1}$.

\subsection{Illustration}

\begin{figure}
\begin{center}
\resizebox{\hsize}{!}{\includegraphics{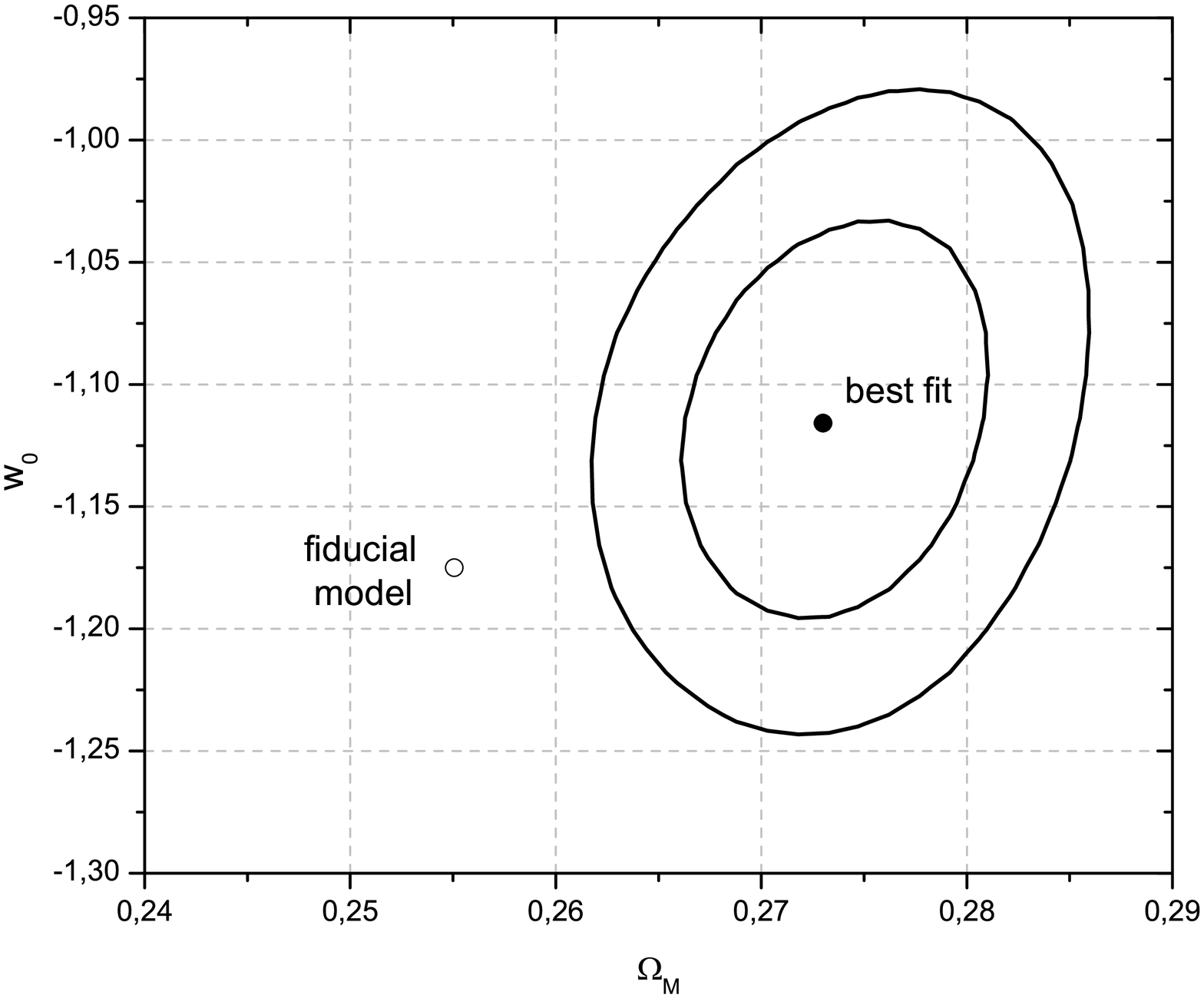}}
\caption{\label{fig:problemillustration} Illustration of the possible risks introduced by the wrong assumption of nonevolving SN magnitudes. The circle depicts the fiducial cosmology $\{\Omega_{\mathrm{M}},w_0,w_a,\beta_1\}^{\mathrm{Fid}}=\{0.255,-1.20,0,-0.10\}$. The best fit with the standard set of cosmological parameters, thus neglecting SN magnitude evolution, is depicted by the full circle, see text. Black lines are the $68\%$ and $95\%$ confidence-level contours.}
\end{center}
\end{figure}

We observe that, even with this high statistics at redshifts $z>1$ and forecasted small errors on $R$ and $A$, an unaccounted for evolution of SN magnitudes will possibly lead to misinterpreting the fit results. Suppose a fiducial cosmological model $\{\Omega_{\mathrm{M}},w_0,w_{\mathrm{a}},\beta_1\}^{\mathrm{Fid}}=\{0.255,-1.20,0,-0.10\}$. This is a cosmos with a dark energy component with a constant equation of state, $w_{\mathrm{x}}={\mathrm{const.}}=w_0$, where SN magnitudes evolve with redshift according to Eq.(\ref{eq:model1}). This fiducial model is represented by the small circle in the $\left(\Omega_{\mathrm{M}},w_0\right)$ parameter space of Fig. \ref{fig:problemillustration}. A fit on this cosmology's SN magnitudes combined with $R$ and $A$ gives a best fit $\{\Omega_{\mathrm{M}},w_0,w_{\mathrm{a}}\}^{\mathrm{fit}}=\{0.274\pm0.007,-1.12\pm0.09,0.23\pm0.41\}$ with a very low $\chi^2/n=0.2$ when the possibility of an SN magnitude evolution is not taken into account in the fitting procedure. Here, $n$ depicts the number of degrees of freedom of the fit. The best-fit value and the corresponding $68\%$ and $95\%$ confidence level contours are plotted in Fig. \ref{fig:problemillustration}, which highlights that the fit result would lead us to reject the true cosmology at more than the $95\%$ confidence level. We point out that the best fit on this cosmology is consistent with the actual real data best fit on SNe+BAO+CMB+WL \citep{Tilquin}, which pins down the parameter values to $\{\Omega_{\mathrm{M}},w_0,w_{\mathrm{a}}\}^{\mathrm{full\;fit}}\sim\{0.28,-1.1,0.4\}$. Even with a mid-term prospective SN dataset and combination with $R$ and $A$ with forecasted small errors, we would therefore risk misinterpreting our best fit because of a wrong model assumption on one of the probes. We have been unable to find an illustration of a confusion between a fiducial $\mathrm{\Lambda CDM}$ model with an SN magnitude evolution, and the `full fit' model given above. Indeed, we find that $w_{\mathrm{a}}^{\;\mathrm{fit}}\leq w_{\mathrm{a}}^{\;\mathrm{Fid}}\left(=0\right)$ whenever $w_0^{\;\mathrm{Fid}}\leq-1$, whatever sign and model of $\Delta m^{\mathrm{evo}}$.

\indent To avoid this misinterpretation, one of course would like to have the means in hand to possibly detect a wrong model assumption on one of the probes. We introduce three different generic criteria that allow judging the performance of a fit. We then check, for which fiducial cosmologies involving an SN magnitude evolution, parameterized by $\beta$, these criteria are strong enough to detect the wrong model assumption in the fit.

\subsection{\label{sec:statisticalcriteria}Statistical goodness criterion} 
As a first criterion for judging the goodness of the fit we use the $\chi^2$-test. To evaluate the statistical goodness of the fit, we choose the $\chi^2_{\mathrm{cut}}$ of the so-called `p-value' \citep{PDG}: 
\begin{equation}
1-0.68\,(0.95)=\int_{\chi_{\mathrm{cut}}^2}^{\infty}\frac{{x^{n/2-1}e^{-x/2}}}{{2^{n/2}\Gamma(n/2)}}dx
\label{eq:statisticalgoodness}
\end{equation}
as statistically justified cut-off values, and consider all models with $\chi^2 > \chi^2_{\mathrm{cut}}$ as ruled out at $68\%\,\left(95\%\right)$ confidence level, respectively. Here, $n$ is the number of degrees of freedom of the fit. The detectability limits on the fiducial parameter $\beta$ we obtain by virtue of this criterion are tabulated in Table \ref{tab:detectabilitylimits}. We furthermore test for the detectability of an evolutionary effect on the SN magnitudes by a consistency criterion which is genuine to SN data.

\subsection{\label{sec:consistency}Consistency Criterion} 
\begin{figure}[h!]
\begin{center}
\resizebox{\hsize}{!}{\includegraphics{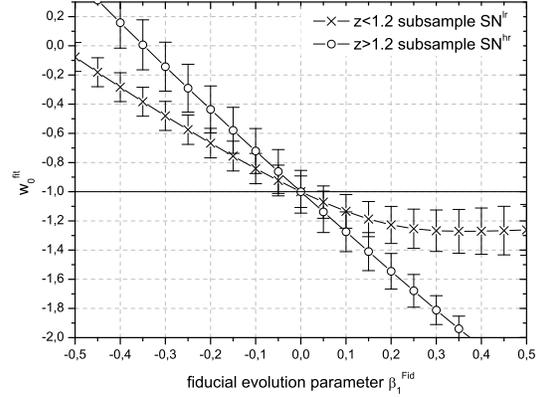}}
\caption{\label{fig:w0consistency} Illustration of the consistency criterion. We plot the $w_0^{\;\mathrm{fit}}$-values obtained for the low redshift $z=0...1.2$ (black crosses) and the high redshift $z=1.2...1.7$  (blue circles) subsample over the fiducial evolution parameter $\beta_1^{\;\mathrm{Fid}}$ of Model $1$. Error bars are the $1\sigma$ errors $\sigma_{w_0}^{\;\;{\mathrm{lr}}}$ and $\sigma_{w_0}^{\;\;{\mathrm{hr}}}$.}
\end{center}
\end{figure}

In the literature it is a common approach to consider the terms `low redshift' and `high redshift' when discussing SN data, where the definition of the limiting redshift between the two ranges, however, is not uniform. Some authors refer to the scatter of the $\Delta m$ in the Hubble diagram Fig. \ref{fig:unionbinned}, and divide the sample into high and low redshift subsamples with equal mean scatter, while others cut the sample by simply assuming that $z>0.5$ or $z>1$ be high redshift. We find it a good statistical means to cut an SN sample into high and low redshift subsamples at a certain $z_{\mathrm{cut}}$ such that the errors on the obtained fit parameters $p_k$ be equal for the two subsamples. That is to say, we impose $\sigma_{\mathrm{p}_k}^{\;\;{\mathrm{hr}}}=\sigma_{\mathrm{p}_k}^{\;\;{\mathrm{lr}}}$ on the high (hr) and low (lr) redshift subsamples. We therewith ensure that the two subsamples have equal statistical weight with respect to the parameter $p_k$, but will find different cut redshifts for different $p_k$. For our simulated SN sample as outlined above, we find $z_{\mathrm{cut}}=1.20\,\left(0.95\right)$ with respect to equal errors on $p_k=\Omega_{\mathrm{M}}\,\left(w_0\right)$. These $z_{\mathrm{cut}}$ are found by looking at the SNe only (and not at the combined set SNe+$R$+$A$), because we want it to be a criterion intrinsic to SNe.\\
\indent Figure \ref{fig:w0consistency} illustrates the application of this criterion for a simulated Model $1$ dataset with $-0.5<\beta_1^{\;\mathrm{Fid}}<0.5$. All other fiducial parameters are fixed to $\{\Omega_{\mathrm{M}},w_0,w_{\mathrm{a}}\}^{\mathrm{Fid}}=\{0.30,-1,0\}$ in the simulation, and the SN sample has been split in two at redshift $z_{\mathrm{cut}}=1.20$. The figure shows the $w_0$ obtained by separate fits on the $\mathrm{SN}^{\mathrm{lr}}$+$R$+$A$ and the $\mathrm{SN}^{\mathrm{hr}}$+$R$+$A$ datasets, plotted over the varying $\beta_1^{\;\mathrm{Fid}}$. We recall that the fitted parameters are $\{p_i\}^{\mathrm{fit}}=\{M_{\mathrm{s}},\Omega_{\mathrm{M}},w_0,w_{\mathrm{a}}\}$ as throughout this section.

\noindent We find, as should be expected, consistent results around $\beta_1^{\;\mathrm{Fid}}=0$, \ie \begin{scriptsize}1.\end{scriptsize} the $w_0^{\;{\mathrm{hr}}}$ obtained by a fit on the high-redshift subsample lie well within the errors of the $w_0^{\;{\mathrm{lr}}}$ obtained by a fit on the low-redshift subsample, and vice versa, and \begin{scriptsize}2.\end{scriptsize} both the $w_0^{\;{\mathrm{hr}}}$ and the $w_0^{\;{\mathrm{lr}}}$ are in agreement with the fiducial $w_0^{\;\mathrm{Fid}}=-1$. But the $w_0$ clearly become inconsistent for larger $|\beta_1^{\;\mathrm{Fid}}|$. To give the exact inconsistency bounds we look at the consistency bias $\mathcal{C}_i=|p_i^{\;{\mathrm{hr}}}-p_i^{\;{\mathrm{lr}}}|$ and consider
\begin{equation}
\frac{\mathcal{C}_i}{T\sqrt{(\sigma_{p_i}^{\;\,\mathrm{hr}})^2+(\sigma_{p_i}^{\;\,\mathrm{lr}})^2}}<1
\label{eq:consistency}
\end{equation}
as the criterion of consistency between the low and high redshift subsamples. Here, $T$ is a factor corresponding to the $68\%\,(95\%)$ confidence level. Values for $T$ are tabulated in the literature \citep{PDG} or can be obtained from integrating $1-0.68\,\left(0.95\right)=\int_{-T}^{T}f(t,n)\mathrm{d}t$, where $f$ is the Student's function and $n$ the number of degrees of freedom of the fit. 
For the specific example of Fig. \ref{fig:w0consistency} we find from Eq.(\ref{eq:consistency}): $-0.17\,\left(-0.40\right)<\beta_1<0.13\,\left(0.23\right)$ as the fiducial parameter range where an SNe+$R$+$A$ analysis gives consistent $w_0$-fit-values on the high- and low-redshift SN subsample. All limits on all models obtained by this criterion are tabulated in Table \ref{tab:detectabilitylimits}. This table also compares the results for $z_{\mathrm{cut}}=1.20$ with those for $z_{\mathrm{cut}}=0.95$.

\subsection{\label{sec:physicalcriteria}Physical criteria} 

We impose
\begin{equation}
(w_0+w_{\mathrm{a}})<0
\label{eq:physicalgoodness1}
\end{equation}
on the fit results. Equation (\ref{eq:physicalgoodness1}) is implied by the condition that $w_{\mathrm{x}}$ be smaller than zero at the time of decoupling, $w_{\mathrm{x}}\left(z_{\mathrm{CMB}}\right)<0$, to allow for a matter-dominated early universe and thus structure formation \citep[see][]{WangSteinhardt}.\\
We also impose
\begin{equation}
\Omega_{\mathrm{M}}>0
\label{eq:physicalgoodness2}
\end{equation}
on the fit to allow for the presence of matter in the universe.\\
It however turns out that in the context of a combined analysis SNe +$R$ +$A$ neither Eq.(\ref{eq:physicalgoodness1}) nor Eq.(\ref{eq:physicalgoodness2}) contribute any constraint on the $\beta_i^{\;\mathrm{Fid}}$ in the range of study. We therefore do not discuss these criteria in the following, and omit them from Table \ref{tab:detectabilitylimits}.

\subsection{\label{sec:results}Results}
\subsubsection{Detectability}

Putting together the criteria developed above, we obtain $\beta$-ranges for all four fiducial evolutionary models where an SN magnitude evolution would pass the fitting procedure undetected by these statistical and physical detectability criteria. Table \ref{tab:detectabilitylimits} shows the ensemble of the results. 
\begin{table*}[h!]
\caption{\label{tab:detectabilitylimits} The $\beta$-intervall in which an evolutionary effect on SN magnitudes would pass undetected by statistical and physical detectability criteria.}
\begin{center}
\begin{tiny}
\begin{tabular}{r|r@{$\,$,$\,$}l|r@{$\,$,$\,$}l|r@{$\,$,$\,$}l|r@{$\,$,$\,$}l||r@{$\,$,$\,$}l|r@{$\,$,$\,$}l|r@{$\,$,$\,$}l|r@{$\,$,$\,$}l}
\hline
\hline
 & \multicolumn{8}{|c||}{\textbf{Model $1$}} & \multicolumn{8}{c}{\textbf{Model $2.1$}} \\
 & \multicolumn{8}{|c||}{} & \multicolumn{8}{c}{} \\
\multicolumn{1}{c|}{Criterion} & \multicolumn{4}{c|}{$68\%$ CL} & \multicolumn{4}{c||}{$95\%$ CL} & \multicolumn{4}{c|}{$68\%$ CL} & \multicolumn{4}{c}{$95\%$ CL} \\
\hline
$\chi^2<\chi^2_{cut}$ & \multicolumn{4}{c|}{$[\;-0.24,0.26\;]$} & \multicolumn{4}{c||}{$[\;\underline{-0.28},0.32\;]$} & \multicolumn{4}{c|}{$[\;-0.28,0.28\;]$} & \multicolumn{4}{c}{$[\;\underline{-0.33},0.33\;]$} \\

 & \multicolumn{4}{c|}{} & \multicolumn{4}{c||}{} & \multicolumn{4}{c|}{} & \multicolumn{4}{c}{} \\
  
\multicolumn{1}{c|}{Consistency} & \multicolumn{2}{c|}{$z_{cut}=0.95$} & \multicolumn{2}{c|}{$z_{cut}=1.20$} & \multicolumn{2}{c|}{$z_{cut}=0.95$} & \multicolumn{2}{c||}{$z_{cut}=1.20$} & \multicolumn{2}{c|}{$z_{cut}=0.95$} & \multicolumn{2}{c|}{$z_{cut}=1.20$} & \multicolumn{2}{c|}{$z_{cut}=0.95$} & \multicolumn{2}{c}{$z_{cut}=1.20$} \\
\multicolumn{1}{c|}{of:} & \multicolumn{2}{c|}{} & \multicolumn{2}{c|}{} & \multicolumn{2}{c|}{} & \multicolumn{2}{c||}{} & \multicolumn{2}{c|}{} & \multicolumn{2}{c|}{} & \multicolumn{2}{c|}{} & \multicolumn{2}{c}{} \\
\multicolumn{1}{c|}{$\Omega_{\mathrm{M}}^{lr} \leftrightarrow \Omega_{\mathrm{M}}^{hr}$} & $[\;-0.20$ & $0.18\;]$ & $[\;-0.18$ & $0.15\;]$ & $[\;(-0.45)$ & $0.38\;]$ & $[\;-0.43$ & $0.28\;]$ & $[\;-0.23$ & $0.20\;]$ & $[\;-0.20$ & $0.17\;]$ & $[\,(-0.6)$ & $0.37\;]$ & $[\;-0.45$ & $0.32\;]$ \\
\multicolumn{1}{c|}{$w_0^{lr} \leftrightarrow w_0^{hr}$}  & $[\;-0.39$ & $0.33\;]$ & $[\;\underline{-0.17}$ & $\underline{0.13}\;]$ & \multicolumn{2}{c|}{$(...)$} & $[\;-0.40$ & $\underline{0.23}\;]$ & $[\,(-0.6)$ & $0.35\;]$ & $[\;\underline{-0.19}$ & $\underline{0.15}\;]$ & $[\,\,\:\:(...)\;\,$ & $0.44\;]$ & $[\;-0.44$ & $\underline{0.26}\;]$ \\
\multicolumn{1}{c|}{$w_a^{lr} \leftrightarrow w_a^{hr}$} & $[\,\,\:\:(...)\;\,$ & $0.41\;]$ & $[\;-0.22$ & $0.15\;]$ & \multicolumn{2}{c|}{$(...)$} & $[\,\,\:\:(...)\;\,$ & $0.26\;]$ & $[\,\,\:\:(...)\;\,$ & $0.42\;]$ & $[\;-0.25$ & $0.17\;]$ & \multicolumn{2}{c|}{$(...)$} & $[\,\,\:\:(...)\;\,$ & $0.30\;]$ \\
\hline
\multicolumn{17}{c}{} \\
\multicolumn{17}{c}{} \\
\hline
\hline
 & \multicolumn{8}{|c||}{\textbf{Model $2.2$}} & \multicolumn{8}{c}{\textbf{Model $2.3$}} \\
 & \multicolumn{8}{|c||}{} & \multicolumn{8}{c}{} \\
\multicolumn{1}{c|}{Criterion} & \multicolumn{4}{c|}{$68\%$ CL} & \multicolumn{4}{c||}{$95\%$ CL} & \multicolumn{4}{c|}{$68\%$ CL} & \multicolumn{4}{c}{$95\%$ CL} \\
\hline
$\chi^2<\chi^2_{cut}$ & \multicolumn{4}{c|}{$[\;-0.12,0.14\;]$} & \multicolumn{4}{c||}{$[\;\underline{-0.14},0.16\;]$} & \multicolumn{4}{c|}{$[\;-0.05,0.05\;]$} & \multicolumn{4}{c}{$[\;\underline{-0.06},\underline{0.06}\;]$} \\

 & \multicolumn{4}{c|}{} & \multicolumn{4}{c||}{} & \multicolumn{4}{c|}{} & \multicolumn{4}{c}{} \\
  
\multicolumn{1}{c|}{Consistency} & \multicolumn{2}{c|}{$z_{cut}=0.95$} & \multicolumn{2}{c|}{$z_{cut}=1.20$} & \multicolumn{2}{c|}{$z_{cut}=0.95$} & \multicolumn{2}{c||}{$z_{cut}=1.20$} & \multicolumn{2}{c|}{$z_{cut}=0.95$} & \multicolumn{2}{c|}{$z_{cut}=1.20$} & \multicolumn{2}{c|}{$z_{cut}=0.95$} & \multicolumn{2}{c}{$z_{cut}=1.20$} \\
\multicolumn{1}{c|}{of:} & \multicolumn{2}{c|}{} & \multicolumn{2}{c|}{} & \multicolumn{2}{c|}{} & \multicolumn{2}{c||}{} & \multicolumn{2}{c|}{} & \multicolumn{2}{c|}{} & \multicolumn{2}{c|}{} & \multicolumn{2}{c}{} \\
\multicolumn{1}{c|}{$\Omega_{\mathrm{M}}^{lr} \leftrightarrow \Omega_{\mathrm{M}}^{hr}$} & $[\;-0.10$ & $0.11\;]$ & $[\;\underline{-0.09}$ & $0.08\;]$ & $[\;-0.21$ & $0.23\;]$ & $[\;-0.21$ & $0.14\;]$ & $[\;-0.04$ & $0.06\;]$ & $[\;\underline{-0.04}$ & $\underline{0.04}\;]$ & $[\;-0.08$ & $0.12\;]$ & $[\;-0.08$ & $0.07\;]$ \\
\multicolumn{1}{c|}{$w_0^{lr} \leftrightarrow w_0^{hr}$}  & $[\;-0.20$ & $0.26\;]$ & $[\;\underline{-0.09}$ & $\underline{0.07}\;]$ & $[\;-0.38$ & $0.43\;]$ & $[\;-0.21$ & $\underline{0.13}\;]$ & $[\;-0.09$ & $0.20\;]$ & $[\;\underline{-0.04}$ & $\underline{0.04}\;]$ & $[\;-0.16$ & $0.28\;]$ & $[\;-0.09$ & $0.07\;]$ \\

\multicolumn{1}{c|}{$w_a^{lr} \leftrightarrow w_a^{hr}$} & $[\,(-0.6)$ & $0.36\;]$ & $[\;-0.12$ & $0.08\;]$ & $[\,\,\:\:(...)\;\,$ & $(0.7)\;]$ & $[\,(-0.3)$ & $0.15\;]$ & $[\;-0.24$ & $0.28\;]$ & $[\;-0.07$ & $0.07\;]$ & $[\,(-0.4)$ & $(0.4)\,]$ & $[\;-0.15$ & $0.09\;]$ \\
\hline
\end{tabular}
\end{tiny}
\end{center}
Values in round brackets $()$ lie outside the scan range, and were obtained by extrapolation. $(...)$ means that no limits fell within the scan range and no meaningful extrapolation was possible.
\end{table*}
The consistency check is particularly powerful, and on the $68\%$ confidence level yet more decisive than the $\chi^2$-test in all cases. At the $95\%$ confidence level, the $\chi^2$-check performs better for $\beta<0$ for all models. We note that the consistency check for $\Omega_{\mathrm{M}}$ is most powerful when the sample is split in two with respect to $\sigma_{w_0}$, \ie when $z_{\mathrm{cut}}=0.95$. Inversely, when the cut is done with respect to $\sigma_{\Omega_{\mathrm{M}}}$, \ie at $z_{\mathrm{cut}}=1.20$, the most constraining test is the consistency check on $w_0$. We underline the most constraining detectablitity limits in Table \ref{tab:detectabilitylimits}. We can translate the underlined $68\%$ confidence level $\beta$-limits from Table \ref{tab:detectabilitylimits} into magnitude shifts ${\Delta m^{\mathrm{evo}}}_{\vert1.7}$ at redshift $z=1.7$ via Eq. (\ref{eq:model1}) and (\ref{eq:models2}), and obtain the Table \ref{tab:nondetectability}. 
\begin{center}
\begin{table}[h!]
\caption{\label{tab:nondetectability} The ${\Delta m^{\mathrm{evo}}}_{\vert1.7}$-intervall in which an SN magnitude evolution would not be detectable at the $68\%$ confidence level.}
\begin{tabular*}{\columnwidth}{@{\extracolsep{\fill}}r|cc}
\hline
\hline
Model  & Non-Detectability Zone & \\
\hline
$1$   & $[\;-0.18,0.14\;]$ & \\
$2.1$ & $[\;-0.25,0.20\;]$ & \\
$2.2$ & $[\;-0.15,0.12\;]$ & \\
$2.3$ & $[\;-0.12,0.12\;]$ & \\

\hline
\end{tabular*}
\end{table}
\end{center}
We highlight that, at the $68\%$ confidence level, the most performing detection criterion is the consistency check. However, as noted, when going to higher confidence levels, one may find the $\chi^2$-check better suited to detecting the effect of SN magnitude evolution. From these $\Delta m^{\mathrm{evo}}$ limits one concludes that the power models Eq.(\ref{eq:models2}) are the higher constraint, the larger the exponent $\alpha$ is. The early-epoch model $2.3$ is detectable whenever the evolutionary magnitude shift $|{\Delta m^{\mathrm{evo}}}_{\vert1.7}|>0.12$ at redshift $1.7$. The late-epoch model $2.1$ would pass undetected for all evolutionary effects $|{\Delta m^{\mathrm{evo}}}_{\vert1.7}|<0.20$ even in an SN survey with redshifts up to $1.7$ and combination with $R$ and $A$. The detectability limits derived in this section also underline the tendency that a negative magnitude drift $\Delta m^{\mathrm{evo}}<0$ is harder to detect than a positive one. We saw the preference for $\Delta m^{\mathrm{evo}}<0$ from real data in section \ref{sec:realdata}, cf. Figs. \ref{fig:bestfitillustration} and \ref{fig:unionbinned}.

\subsubsection{Biases}

Whether this undetectability of the SN magnitude evolution is dangerous or not depends on the quality of reconstruction of the fit parameters $\{p_i\}$. For a fiducial cosmology involving, \eg, a $\beta_1=-0.03$ (corresponding to ${\Delta m^{\mathrm{evo}}}_{\vert1.7}=-0.03$) evolutionary effect on SN magnitudes, the problem would not be detected. But we would not need to worry because all the fiducial parameters $\{p_i\}^{\;\mathrm{Fid}}=\{\Omega_{\mathrm{M}},w_0,w_{\mathrm{a}}\}^{\mathrm{Fid}}$ lie within the $1\sigma$ error of the fitted value, and the fiducial cosmology would thus be `correctly' reconstructed at the $1\sigma$ level despite the undetected SN magnitude evolution. We introduce the bias $\mathcal{B}_i$ on parameter $p_i$ as $\mathcal{B}_i=|p_i^{\;\mathrm{fit}}-p_i^{\;\mathrm{Fid}}|$, and adopt the notion that the parameter $p_i$ is `biased' whenever:
\begin{equation}
\frac{\mathcal{B}_i}{\sigma_{p_i}}>1.
\label{eq:bias}
\end{equation}
By applying this notion, in this section we restrict ourselves to the $68\%$ confidence level. We call the zones of fiducial parameter space where parameters $p_k$ are reconstructed with biases the bias zones. Correspondingly, the zones of correct reconstruction without bias are referred to as validity zones.\\ 
Whereas no bias is introduced in the fit in the preceding example, fiducial models exist where an evolutionary effect $\Delta m^{\mathrm{evo}}$ would pass our criteria undetected, and where, at the same time, one or several of the $\{p_i\}$ are biased. Our simulations, \eg, show that in a universe with an unaccounted for Model $1$ like SN magnitude evolution, the $\Omega_{\mathrm{M}}$ parameter is biased whenever $|{\Delta m^{\mathrm{evo}}}_{\vert1.7}|\geq0.04$. We concluded in the previous section that only ${\Delta m^{\mathrm{evo}}}_{\vert1.7}<-0.18$ and ${\Delta m^{\mathrm{evo}}}_{\vert1.7}>0.14$ can be detected, cf. Table \ref{tab:detectabilitylimits}. Our ignorance of the evolution thus leaves us with a dangerous zone where an SN magnitude evolution undetectably biases the results. A full study of all the $\{p_i\}$ for models $1$ and $2$ yields the results of Table \ref{tab:biaslimits}, which gives the validity zones for $\Omega_{\mathrm{M}}$, $w_0$ and $w_{\mathrm{a}}$, for the four models. 
\begin{center}
\begin{table}[h!]
\caption{\label{tab:biaslimits} The ${\Delta m^{\mathrm{evo}}}_{\vert1.7}$-intervall in which the fit parameters are reconstructed without biases.}
\begin{tabular*}{\columnwidth}{@{\extracolsep{\fill}}r|ccc}
\hline
\hline
& \multicolumn{3}{c}{Validity Zone}\\
Model & $\Omega_{\mathrm{M}}$ & $w_0$ & $w_{\mathrm{a}}$ \\
\hline
$1$      & $[\;-0.04,0.04\;]$ & $[\;-0.08,1.18\;]$ & $[\;-0.28,0.15\;]$ \\
$2.1$    & $[\;-0.04,0.04\;]$ & $[\;-0.08,0.13\;]$ & $[\;-0.27,0.27\;]$ \\
$2.2$    & $[\;-0.03,0.05\;]$ & $[\;-0.10,0.37\;]$ & $[\;-0.39,0.09\;]$ \\
$2.3$    & $[\;-0.06,0.09\;]$ & $[\;-0.17,0.09\;]$ & $[\;-0.06,0.06\;]$ \\
\hline
\end{tabular*}
\end{table}
\end{center}
We see that the parameter $\Omega_{\mathrm{M}}$ has the highest chances of being biased for all models. Its validity zones are the smallest in comparison to the $w_0$ and $w_{\mathrm{a}}$ zones. All parameters $\{\Omega_{\mathrm{M}},w_0,w_{\mathrm{a}}\}$, however, share the general tendency: the higher the power $\alpha$ of models $2$, the larger the validity zone. A particular feature of model $1$ is that $w_0$ is reconstructed without bias for all positive fiducial ${\Delta m^{\mathrm{evo}}}_{\vert1.7}<1.18$.

\subsubsection{Danger}

\begin{figure*}[h!]
\begin{center}
\includegraphics[width=17cm]{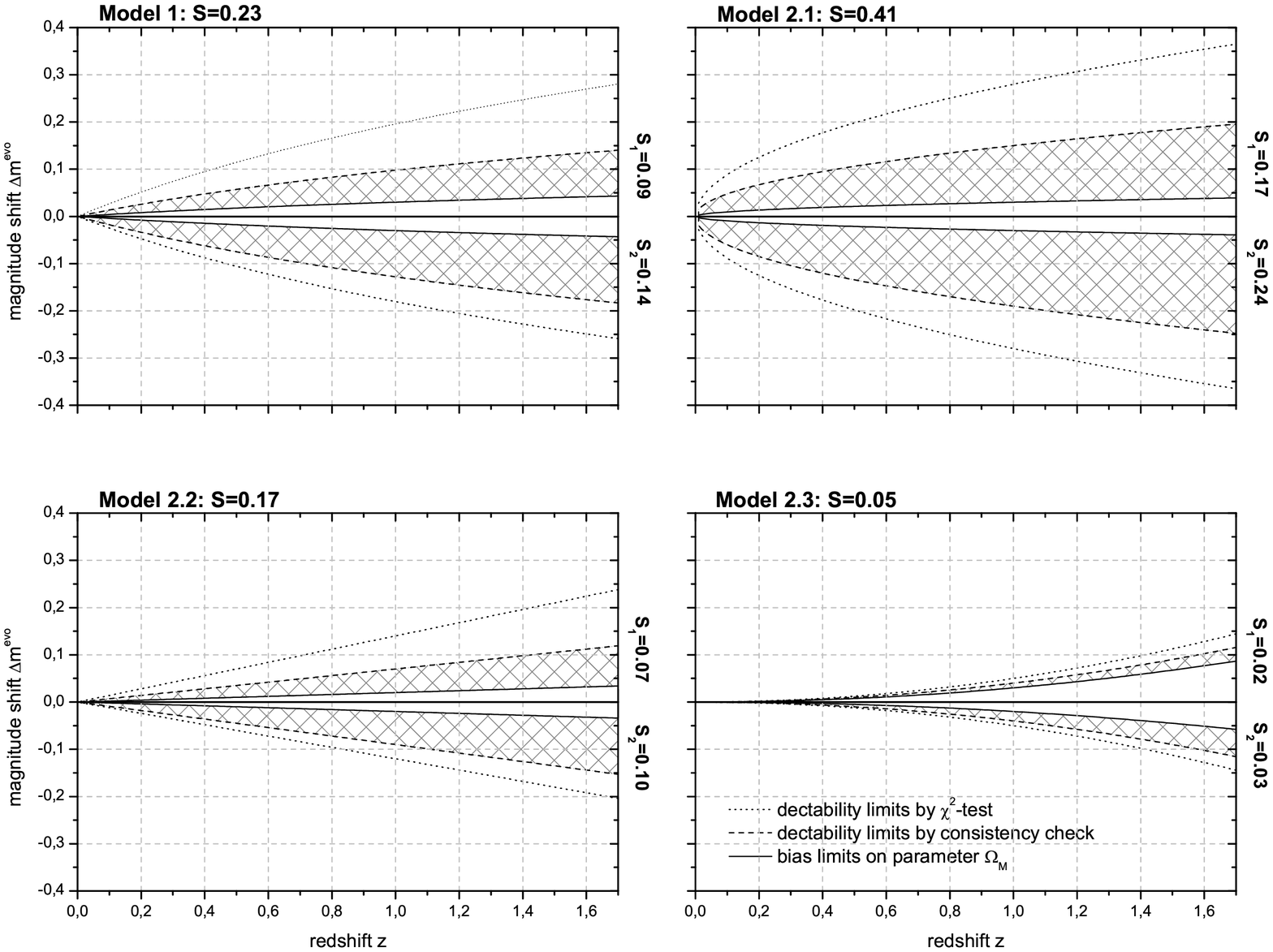}
\caption{\label{fig:alldangerousness} $68\%$ confidence level detectability and bias limits, and the dangerous zones in between (cross-shaded region).}
\end{center}
\end{figure*}

We can merge the obtained detectability and bias-risk limits, taken from Tables \ref{tab:detectabilitylimits} and \ref{tab:biaslimits}, respectively, to estimate the danger of a neglected magnitude evolution governed by the models of Eqs. (\ref{eq:model1}) and (\ref{eq:models2}). We limit our discussion to the parameter $\Omega_{\mathrm{M}}$, because it is the parameter with the highest risk of undetected biased reconstruction for all models. Figure \ref{fig:alldangerousness} shows the $68\%$ confidence level detectability and bias limits for the four models obtained from Tabels \ref{tab:detectabilitylimits} and \ref{tab:biaslimits} in a magnitude-redshift-diagram. 
Dotted lines are the detectability limits obtained by the $\chi^2$-test, dashed lines are the detectability limits obtained by the most constraining consistency check (underlined values in table \ref{tab:detectabilitylimits}), and full lines are the bias limits on $\Omega_{\mathrm{M}}$ taken from Table \ref{tab:biaslimits}.\\
The cross-shaded zones between the detectability and the bias limits are the dangerous regions of the diagram, where the fiducial evolutionary effect passes the analysis undetected, and where biases are introduced  at the same time on the parameter $\Omega_{\mathrm{M}}$. We may take the area $S$ of the cross-shaded region, in the redshift range $0<z<1.7$, as a measure of the danger of the model. The higher the value of $S$, the greater the risks of misinterpretation. We obtain the values of Table \ref{tab:dangerousness}. 
\begin{center}
\begin{table}[h!]
\caption{\label{tab:dangerousness} Danger $S$ of the four models as described in the text.}
\begin{tabular*}{\columnwidth}{@{\extracolsep{\fill}}r|ccc}
\hline
\hline
Model & $\;\;{S_1}^{*}\;\;$ & $\;\;{S_2}^{*}\;\;$ & $\;\;S\;\;$ \\
\hline
$1$   & $0.09$ & $0.14$ & $0.23$ \\
$2.1$ & $0.17$ & $0.24$ & $0.41$ \\
$2.2$ & $0.07$ & $0.10$ & $0.17$ \\
$2.3$ & $0.02$ & $0.03$ & $0.05$ \\
\hline
\end{tabular*}
$^*\;$Size of the dangerous zone in the upper (lower) part of the Hubble diagram only, cf. Fig. \ref{fig:alldangerousness}.
\end{table}
\end{center}
These values validate the tendency of higher power $\alpha\ga1$ evolutionary effects being less dangerous than $\alpha\la1$ models. The late-epoch model $2.1$ is more likely to undetectably bias the fit results than the early-epoch model $2.3$. The danger of the linear model $2.2$ and the logarithmic model $1$ lie in between the two extremes, having approximately the same values. Figure \ref{fig:alldangerousness} also shows the values $S_1\,\left(S_2\right)$ of the area of the upper (lower) dangerous region seperately. One finds $S_2>S_1$ for all models, which again means that a negative magnitude drift $\Delta m^{\mathrm{evo}}<0$ carries the greater risk of undetectably biasing the fit results than a positive one.\\
We also extracted the danger of the models used by \citet{Linder2006a} and \citet{Ferramacho}: $S=0.36$, which turn out to be equivalent, and very close to our Model $2.1$. This agrees with these parameterizations being well described by a model $2$ with $\alpha \approx 0.6$.\\
The analysis of the $95\%$ confidence level naturally yields higher absolute values for $S$, but the same conclusions hold. In particular, we obtain the same tendencies within the models, and $S_2>S_1$ for all models.\\
The absolute value of $S$ also depends on the specifications of the data survey, \ie in our case on errors on $R$ and $A$, and on the statistics of the SNe in the redshift range. For example, when we neglect the assumed systematic error $\sigma_{\mathrm{sys}}=0.02$ and decrease the intrinsic magnitude error on each SN to $0.10$, we obtain bias limits that are tighter than those of Table \ref{tab:biaslimits} roughly by a factor $2$. Also the dectectability limits from Table \ref{tab:detectabilitylimits} decrease considerably from 30\% to 50\%, and the $\chi^2$ test becomes more and more effective the smaller the intrinsic errors on magnitudes. The absolute values of $S$ also depend on the priors of the fit. They may change when the flatness constraint is dropped or when $w_{\mathrm{x}}=\mathrm{const.}$ is imposed on the fit. In the latter case, we find a decrease in the detectability ranges and also the bias limits. In particular, when imposing $w_{\mathrm{x}}=\mathrm{const.}$ one obtains results consistent with those of \citet{SarkarSubpopulations}, \ie $\Omega_{\mathrm{M}}$ and $w_{\mathrm{x}}$ are biased at $1\sigma$ for evolutionary effects ${\Delta m^{\mathrm{evo}}}_{\vert1.7}=0.02$ even in the Model $1$ case.\\
The measure of danger $S$ we introduce in this section therefore should not be taken as an absolute measure, because its actual values depend on the specifications of the cosmological probes applied in the fit, and on priors. However, we checked that variations in the data model or of the priors do not change the conclusion that the late-epoch model $\Delta m^{\mathrm{evo}}$ is the most dangerous one.

\section{\label{sec:discussion}Discussion}

We studied four different one-parametric models of SN magnitude evolution on cosmic time scales and obtained constraints on its parameters by combined fits on the actual real data coming from Supernova surveys, observations of the cosmic microwave background, and baryonic acoustic oscillation. We found by a minimization of the $\chi^2$ that data prefer a magnitude evolution of SNe type Ia such that high-redshift supernovae are brighter than would be expected in a standard $\{\Omega_i,w_{\mathrm{x}}\}$ cosmos: $\Delta m^{\mathrm{evo}}<0$. Data are, however, consistent with nonevolving magnitudes at the $1\sigma$ level except for special cases. The special cases we found are a fit with the early-epoch model $2.3$ when dark energy is allowed to be dynamical, $w_{\mathrm{x}}=w_0+w_{\mathrm{a}}z/\left(1+z\right)$, and a fit with a logarithmic magnitude evolution Model $1$ when a constant equation of state of dark energy $w_{\mathrm{x}}=\mathrm{const.}$ is assumed in the fit. A comparison of the SN magnitude distribution with the CPL best-fit magnitudes in a Hubble-diagram also indicate more luminous SN events at higher redshift. Our results are, however, of limited strength because we do not use the full CMB and BAO data but the reduced variables $R$ and $A$. We also did not include constraints coming from other cosmological probes.\\ 
We simulated a future data scenario consisting of $2000$ SN events out to redshift $z=1.7$ and forecasted small errors on the CMB shift parameter $R$ and the BAO variable $A$. In the simulation of the SN magnitudes, a redshift dependence of the magnitudes according to the four models was allowed for. Then in the fit we neglected the possibility of such an evolution and studied the fit results with respect to detectability of this wrong model assumption for a wide range of fiducial models. We quantified the range of values of the fiducial evolution parameters $\beta$ where the wrong assumption is not detectable. We found that the linear model $2.2$ and the early-epoch model $2.3$ are easier to detect and reject from the fit than is the late-epoch model $2.1$ and the logarithmic model $1$.\\
\indent Including the possible biasing of the fitted parameters when neglecting magnitude evolution in the discussion, we were able to determine the danger of the various fiducial evolutionary models; that is to say, we determined the exact parameter zones where the wrong model assumption of nonevolving intrinsic SN magnitude not only is not detectable, but also introduces biases on the fitted cosmological parameters $\{p_i\}$. The parameter of overall mass density $\Omega_{\mathrm{M}}$ turned out to carry the highest risk of biased reconstruction. We found that, whereas the dangerous zone is nearly negligible for the early-epoch model $2.3$ ($\Delta m^{\mathrm{evo}}\sim z^2$), it becomes significant in the logarithmic model $1$ ($\Delta m^{\mathrm{evo}}\sim \log{\left(1+z\right)}$) and the linear model $2.2$ ($\Delta m^{\mathrm{evo}}\sim z$), and dangerous for the late epoch model $2.1$ ($\Delta m^{\mathrm{evo}}\sim \sqrt{z}$). This becomes apparent in Fig. \ref{fig:alldangerousness}, where we simultaneously plotted the detectability and bias limits of the four fiducial models in a Hubble diagram. A comparison of Figs. \ref{fig:alldangerousness}, \ref{fig:bestfitillustration}, and \ref{fig:unionbinned} shows that the dangerous zone of Model $2.1$ covers nearly the whole area of magnitude dispersion of actual SN data in the magnitude-redshift-diagram. Also negative magnitude drifts $\left(\Delta m^{\mathrm{evo}}<0\right)$ carry the greater risk of undetectably biasing the fit results than magnitude evolution effects with $\Delta m^{\mathrm{evo}}>0$, which are already excluded at $1\sigma$ by present data for dynamical dark energy.\\

\noindent In conclusion, special care should be accorded to effects yielding late-time evolution of SN Ia magnitudes with a negative magnitude drift. Early-time evolution is less severe. To avoid any bias in determining the cosmological parameters, it is preferable to include a new parameter to describe possible SN Ia magnitude evolution in any combined analysis. From our results, it appears that the models chosen to describe such an effect should favor late-epoch evolution, as is done by our Model 2.1, $\Delta m^{\mathrm{evo}}\sim\sqrt{z}$.

\begin{acknowledgements} We acknowledge useful discussions with A. Blanchard and P.-S. Corasaniti. We thank A. Ealet, C. Tao, C. Marinoni, P. Taxil and D. Fouchez for many interesting discussions. S.L. thanks the \upshape{Gottlieb Daimler- and Karl Benz-Foundation} and the \upshape{DAAD} for financial support.
\end{acknowledgements}

\bibliographystyle{aa}
\bibliography{P039}

\begin{thebibliography}{42}
\expandafter\ifx\csname natexlab\endcsname\relax\def\natexlab#1{#1}\fi

\bibitem[{{Aguirre}(1999)}]{Aguirre}
{Aguirre}, A. 1999, \apj, 525, 583

\bibitem[{Amsler {et~al.}(2008)}]{PDG}
Amsler, C. {et~al.} 2008, Phys. Lett. B, 667, 1

\bibitem[{{Branch} {et~al.}(2001){Branch}, {Perlmutter}, {Baron}, \&
  {Nugent}}]{Branch01}
{Branch}, D., {Perlmutter}, S., {Baron}, E., \& {Nugent}, P. 2001, in The SNAP
  (Supernova Acceleration Probe) Yellow (Snowmass)

\bibitem[{Bronder {et~al.}(2007)}]{SNLS}
Bronder, T.~J. {et~al.} 2007

\bibitem[{{Chevallier} \& {Polarski}(2001)}]{CP}
{Chevallier}, M. \& {Polarski}, D. 2001, Int. J. Mod. Phys. C, 10, 213

\bibitem[{{Corasaniti}(2006)}]{Corasaniti}
{Corasaniti}, P.~S. 2006, \mnras, 372, 191

\bibitem[{Drell {et~al.}(2000)Drell, Loredo, \& Wassermann}]{Drell}
Drell, P.~S., Loredo, T.~J., \& Wassermann, I. 2000, ApJ, 530, 593

\bibitem[{{Efstathiou} \& {Bond}(1999)}]{Efstathiou}
{Efstathiou}, G. \& {Bond}, J.~R. 1999, \mnras, 304, 75

\bibitem[{{Eisenstein} {et~al.}(2005){Eisenstein}, {Zehavi}, {Hogg},
  {Scoccimarro}, {Blanton}, {Nichol}, {Scranton}, {Seo}, {Tegmark}, {Zheng},
  {Anderson}, {Annis}, {Bahcall}, {Brinkmann}, {Burles}, {Castander},
  {Connolly}, {Csabai}, {Doi}, {Fukugita}, {Frieman}, {Glazebrook}, {Gunn},
  {Hendry}, {Hennessy}, {Ivezi{\'c}}, {Kent}, {Knapp}, {Lin}, {Loh}, {Lupton},
  {Margon}, {McKay}, {Meiksin}, {Munn}, {Pope}, {Richmond}, {Schlegel},
  {Schneider}, {Shimasaku}, {Stoughton}, {Strauss}, {SubbaRao}, {Szalay},
  {Szapudi}, {Tucker}, {Yanny}, \& {York}}]{Eisenstein05}
{Eisenstein}, D.~J., {Zehavi}, I., {Hogg}, D.~W., {et~al.} 2005, \apj, 633, 560

\bibitem[{{Ferramacho} {et~al.}(2009){Ferramacho}, {Blanchard}, \&
  {Zolnierowski}}]{Ferramacho}
{Ferramacho}, L.~D., {Blanchard}, A., \& {Zolnierowski}, Y. 2009, \aap, 499, 21

\bibitem[{{Gallagher} {et~al.}(2005){Gallagher}, {Garnavich}, {Berlind},
  {Challis}, {Jha}, \& {Kirshner}}]{Gallagher}
{Gallagher}, J.~S., {Garnavich}, P.~M., {Berlind}, P., {et~al.} 2005, \apj,
  634, 210

\bibitem[{{Guy} {et~al.}(2007){Guy}, {Astier}, {Baumont}, {Hardin}, {Pain},
  {Regnault}, {Basa}, {Carlberg}, {Conley}, {Fabbro}, {Fouchez}, {Hook},
  {Howell}, {Perrett}, {Pritchet}, {Rich}, {Sullivan}, {Antilogus}, {Aubourg},
  {Bazin}, {Bronder}, {Filiol}, {Palanque-Delabrouille}, {Ripoche}, \&
  {Ruhlmann-Kleider}}]{Guy}
{Guy}, J., {Astier}, P., {Baumont}, S., {et~al.} 2007, \aap, 466, 11

\bibitem[{{Hamuy} {et~al.}(1996{\natexlab{a}}){Hamuy}, {Phillips}, {Suntzeff},
  {Schommer}, {Maza}, \& {Aviles}}]{Hamuy961}
{Hamuy}, M., {Phillips}, M.~M., {Suntzeff}, N.~B., {et~al.} 1996{\natexlab{a}},
  \aj, 112, 2391

\bibitem[{{Hamuy} {et~al.}(1996{\natexlab{b}}){Hamuy}, {Phillips}, {Suntzeff},
  {Schommer}, {Maza}, \& {Aviles}}]{Hamuy962}
{Hamuy}, M., {Phillips}, M.~M., {Suntzeff}, N.~B., {et~al.} 1996{\natexlab{b}},
  \aj, 112, 2398

\bibitem[{{Hatano} {et~al.}(2000){Hatano}, {Branch}, {Lentz}, {Baron},
  {Filippenko}, \& {Garnavich}}]{Hatano}
{Hatano}, K., {Branch}, D., {Lentz}, E.~J., {et~al.} 2000, \apjl, 543, L49

\bibitem[{{Hoeflich} {et~al.}(1998){Hoeflich}, {Wheeler}, \&
  {Thielemann}}]{Hoeflich}
{Hoeflich}, P., {Wheeler}, J.~C., \& {Thielemann}, F.~K. 1998, \apj, 495, 617

\bibitem[{{Howell} {et~al.}(2007){Howell}, {Sullivan}, {Conley}, \&
  {Carlberg}}]{Howell07}
{Howell}, D.~A., {Sullivan}, M., {Conley}, A., \& {Carlberg}, R. 2007, \apjl,
  667, L37

\bibitem[{{Hoyle} \& {Fowler}(1960)}]{Hoyle60}
{Hoyle}, F. \& {Fowler}, W.~A. 1960, \apj, 132, 565

\bibitem[{{Kim} {et~al.}(2004){Kim}, {Linder}, {Miquel}, \& {Mostek}}]{Kim}
{Kim}, A.~G., {Linder}, E.~V., {Miquel}, R., \& {Mostek}, N. 2004, \mnras, 347,
  909

\bibitem[{{Komatsu} {et~al.}(2009){Komatsu}, {Dunkley}, {Nolta}, {Bennett},
  {Gold}, {Hinshaw}, {Jarosik}, {Larson}, {Limon}, {Page}, {Spergel},
  {Halpern}, {Hill}, {Kogut}, {Meyer}, {Tucker}, {Weiland}, {Wollack}, \&
  {Wright}}]{Komatsu}
{Komatsu}, E., {Dunkley}, J., {Nolta}, M.~R., {et~al.} 2009, \apjs, 180, 330

\bibitem[{{Kowalski} {et~al.}(2008){Kowalski}, {Rubin}, {Aldering},
  {Agostinho}, {Amadon}, {Amanullah}, {Balland}, {Barbary}, {Blanc}, {Challis},
  {Conley}, {Connolly}, {Covarrubias}, {Dawson}, {Deustua}, {Ellis}, {Fabbro},
  {Fadeyev}, {Fan}, {Farris}, {Folatelli}, {Frye}, {Garavini}, {Gates},
  {Germany}, {Goldhaber}, {Goldman}, {Goobar}, {Groom}, {Haissinski}, {Hardin},
  {Hook}, {Kent}, {Kim}, {Knop}, {Lidman}, {Linder}, {Mendez}, {Meyers},
  {Miller}, {Moniez}, {Mour{\~a}o}, {Newberg}, {Nobili}, {Nugent}, {Pain},
  {Perdereau}, {Perlmutter}, {Phillips}, {Prasad}, {Quimby}, {Regnault},
  {Rich}, {Rubenstein}, {Ruiz-Lapuente}, {Santos}, {Schaefer}, {Schommer},
  {Smith}, {Soderberg}, {Spadafora}, {Strolger}, {Strovink}, {Suntzeff},
  {Suzuki}, {Thomas}, {Walton}, {Wang}, {Wood-Vasey}, \& {Yun}}]{Kowalski}
{Kowalski}, M., {Rubin}, D., {Aldering}, G., {et~al.} 2008, \apj, 686, 749

\bibitem[{{Lentz} {et~al.}(2000){Lentz}, {Baron}, {Branch}, {Hauschildt}, \&
  {Nugent}}]{Lentz}
{Lentz}, E.~J., {Baron}, E., {Branch}, D., {Hauschildt}, P.~H., \& {Nugent},
  P.~E. 2000, \apj, 530, 966

\bibitem[{{Linder}(2003)}]{L}
{Linder}, E.~V. 2003, \prl, 90, 091301

\bibitem[{{Linder}(2006)}]{Linder2006a}
{Linder}, E.~V. 2006, Astropart. Phys., 26, 102

\bibitem[{{Linder}(2009)}]{Linder09}
{Linder}, E.~V. 2009, \prd, 79, 023509

\bibitem[{{M{\'e}nard} {et~al.}(2009){M{\'e}nard}, {Kilbinger}, \&
  {Scranton}}]{Menard2009}
{M{\'e}nard}, B., {Kilbinger}, M., \& {Scranton}, R. 2009, MNRAS, submitted

\bibitem[{{Nordin} {et~al.}(2008){Nordin}, {Goobar}, \& {J{\"o}nsson}}]{Nordin}
{Nordin}, J., {Goobar}, A., \& {J{\"o}nsson}, J. 2008, JCAP, 2, 8

\bibitem[{{Perlmutter} {et~al.}(1999){Perlmutter}, {Aldering}, {Goldhaber},
  {Knop}, {Nugent}, {Castro}, {Deustua}, {Fabbro}, {Goobar}, {Groom}, {Hook},
  {Kim}, {Kim}, {Lee}, {Nunes}, {Pain}, {Pennypacker}, {Quimby}, {Lidman},
  {Ellis}, {Irwin}, {McMahon}, {Ruiz-Lapuente}, {Walton}, {Schaefer}, {Boyle},
  {Filippenko}, {Matheson}, {Fruchter}, {Panagia}, {Newberg}, {Couch}, \& {The
  Supernova Cosmology Project}}]{P99}
{Perlmutter}, S., {Aldering}, G., {Goldhaber}, G., {et~al.} 1999, \apj, 517,
  565

\bibitem[{{Riess}(2000)}]{Riess}
{Riess}, A.~G. 2000, \pasp, 112, 1284

\bibitem[{{Riess} {et~al.}(1998){Riess}, {Filippenko}, {Challis},
  {Clocchiatti}, {Diercks}, {Garnavich}, {Gilliland}, {Hogan}, {Jha},
  {Kirshner}, {Leibundgut}, {Phillips}, {Reiss}, {Schmidt}, {Schommer},
  {Smith}, {Spyromilio}, {Stubbs}, {Suntzeff}, \& {Tonry}}]{Riess98}
{Riess}, A.~G., {Filippenko}, A.~V., {Challis}, P., {et~al.} 1998, \aj, 116,
  1009

\bibitem[{{Riess} {et~al.}(1999){Riess}, {Filippenko}, {Li}, \&
  {Schmidt}}]{Riess99}
{Riess}, A.~G., {Filippenko}, A.~V., {Li}, W., \& {Schmidt}, B.~P. 1999, \aj,
  118, 2668

\bibitem[{{Riess} \& {Livio}(2006)}]{Riess06}
{Riess}, A.~G. \& {Livio}, M. 2006, \apj, 648, 884

\bibitem[{{R{\"o}pke}(2005)}]{Roepke05}
{R{\"o}pke}, F.~K. 2005, \aap, 432, 969

\bibitem[{{R{\"o}pke} \& {Hillebrandt}(2004)}]{Roepke}
{R{\"o}pke}, F.~K. \& {Hillebrandt}, W. 2004, \aap, 420, L1

\bibitem[{{Sarkar} {et~al.}(2008{\natexlab{a}}){Sarkar}, {Amblard}, {Cooray},
  \& {Holz}}]{SarkarSubpopulations}
{Sarkar}, D., {Amblard}, A., {Cooray}, A., \& {Holz}, D.~E. 2008{\natexlab{a}},
  \apjl, 684, L13

\bibitem[{{Sarkar} {et~al.}(2008{\natexlab{b}}){Sarkar}, {Amblard}, {Holz}, \&
  {Cooray}}]{SarkarLensing}
{Sarkar}, D., {Amblard}, A., {Holz}, D.~E., \& {Cooray}, A. 2008{\natexlab{b}},
  \apj, 678, 1

\bibitem[{{Steinhardt} {et~al.}(1999){Steinhardt}, {Wang}, \&
  {Zlatev}}]{WangSteinhardt}
{Steinhardt}, P.~J., {Wang}, L., \& {Zlatev}, I. 1999, \prd, 59, 123504

\bibitem[{{Strolger} {et~al.}(2004){Strolger}, {Riess}, {Dahlen}, {Livio},
  {Panagia}, {Challis}, {Tonry}, {Filippenko}, {Chornock}, {Ferguson},
  {Koekemoer}, {Mobasher}, {Dickinson}, {Giavalisco}, {Casertano}, {Hook},
  {Blondin}, {Leibundgut}, {Nonino}, {Rosati}, {Spinrad}, {Steidel}, {Stern},
  {Garnavich}, {Matheson}, {Grogin}, {Hornschemeier}, {Kretchmer}, {Laidler},
  {Lee}, {Lucas}, {de Mello}, {Moustakas}, {Ravindranath}, {Richardson}, \&
  {Taylor}}]{SNhighz}
{Strolger}, L.-G., {Riess}, A.~G., {Dahlen}, T., {et~al.} 2004, \apj, 613, 200

\bibitem[{{Sullivan} {et~al.}(2009){Sullivan}, {Ellis}, {Howell}, {Riess},
  {Nugent}, \& {Gal-Yam}}]{Sullivan09}
{Sullivan}, M., {Ellis}, R.~S., {Howell}, D.~A., {et~al.} 2009, \apjl, 693, L76

\bibitem[{Tilquin {et~al.}(in preparation)}]{Tilquin}
Tilquin, A. {et~al.} in preparation

\bibitem[{{Wang} \& {Mukherjee}(2007)}]{Wang}
{Wang}, Y. \& {Mukherjee}, P. 2007, \prd, 76, 103533

\bibitem[{{Wood-Vasey} {et~al.}(2004){Wood-Vasey}, {Aldering}, {Lee}, {Loken},
  {Nugent}, {Perlmutter}, {Siegrist}, {Wang}, {Antilogus}, {Astier}, {Hardin},
  {Pain}, {Copin}, {Smadja}, {Gangler}, {Castera}, {Adam}, {Bacon},
  {Lemonnier}, {P{\'e}contal}, {P{\'e}contal}, \& {Kessler}}]{SNFact}
{Wood-Vasey}, W.~M., {Aldering}, G., {Lee}, B.~C., {et~al.} 2004, New Astron.
  Rev., 48, 637

\end{thebibliography}
\end{document}